\documentclass[%
 reprint,
 amsmath,amssymb,
 aps,
 pra,
]{revtex4-1}

\usepackage{graphicx}% Include figure files
\usepackage{dcolumn}% Align table columns on decimal point
\usepackage{bm}% bold math
\usepackage[caption=false]{subfig}
\usepackage{natbib}
\usepackage{hyperref}% add hypertext capabilities
\begin{document}

%\preprint{APS/123-QED}

\title{Micromotion-Enhanced Fast Entangling Gates For Trapped Ion Quantum Computing}% Force line breaks with \\

\author{Alexander K. Ratcliffe}
\affiliation{Department of Quantum Science, RSPE, Australian National University}
\email{alexander.ratcliffe@anu.edu.au}
\author{Lachlan M. Oberg}
\affiliation{Department of Quantum Science, RSPE, Australian National University}
\author{Joseph J. Hope}
\affiliation{Department of Quantum Science, RSPE, Australian National University}

\date{\today}% It is always \today, today,

\begin{abstract}
RF-induced micromotion in trapped ion systems is typically minimised or circumvented to avoid off-resonant couplings for adiabatic processes such as multi-ion gate operations. Non-adiabatic entangling gates (so-called `fast gates') do not require resolution of specific motional sidebands, and are therefore not limited to timescales longer than the trapping period. We find that fast gates designed for micromotion-free environments have significantly reduced fidelity in the presence of micromotion.  We show that when fast gates are designed to account for the RF-induced micromotion, they can, in fact, out-perform fast gates in the absence of micromotion.  The state-dependent force due to the laser induces energy shifts that are amplified by the state-independent forces producing the micromotion.  This enhancement is present for all trapping parameters and is robust to realistic sources of experimental error.  This result paves the way for fast two-qubit entangling gates on scalable 2D architectures, where micromotion is necessarily present on at least one inter-ion axis.
\end{abstract}

\pacs{03.67.Lx}% PACS, the Physics and Astronomy Classification Scheme.

\maketitle

\section{Introduction}

Quantum computing offers the promise of boosting our current computational capabilities, outperforming certain known classical algorithms and allowing tractable simulations of complex quantum systems \cite{Nielsen2010}. To realise this potential, a quantum information processing (QIP) architecture must be able to scale to large numbers of qubits. Many platforms have been proposed as QIP architectures including superconducting qubits \cite{Makhlin1999}, defect centres in diamonds \cite{Neumann2008}, single photons \cite{Knill2001}, NMR \cite{Cory1997}, topological qubits \cite{Nayak2008}, quantum dots \cite{Loss1998}, and spin-spin interactions in silicon donor sites \cite{Kane1998}.  Significant progress towards the requirements for scalable quantum computing has been made on these platforms, demonstrating single- and two-qubit fidelities \cite{Fowler2012,DiVincenzo2000a,Waldherr2014,Veldhorst2014,Dolde2014,Veldhorst2014,Barends2014} above a 98\% fault-tolerant threshold. However, these platforms have been limited either by scalability or the number of high fidelity operations achievable before state decoherence. To date, trapped ions have been a front-runner of QIP architectures, making a number of important demonstrations towards scalable fault-tolerant quantum computing. These have included the deterministic entanglement between 20 atomic qubits \cite{Friis2018}, the use of hybrid quantum-classical computer to find the ground state energy of simple molecules \cite{Hempel2018}, the topological protection and error correction of a qubit state \cite{Nigg2014}, an analog quantum simulation of phase transitions using 53 qubits \cite{Zhang2017}. These are important demonstrations of the control and procedures required for scalable fault-tolerant quantum computing. 

A key limitation in all QIP platforms is the number of independent, high fidelity two-qubit gate operations that can be achieved within the decoherence time of the qubit. Fault-tolerant quantum computation requires that enough of these operations can be conducted with high enough fidelities to correct both for errors in the gate operations and state decoherence. For surface codes, these limits are modest and can enable fault-tolerant computation with fidelities as low as 98\% \cite{Fowler2012}. In this manuscript we use a more stringent threshold of 99.98\% as a guide, achievable using a Bacon-Shor code with $n=10$ \cite{Brooks2013}. Numerous implementations of side-band resolving adiabatic gates have demonstrated fidelities above the 98\% fault-tolerant threshold. For example Bell state fidelities of 99.9\% and above \cite{Ballance2016,Gaebler2016} have been demonstrated, and a gate fidelity of 99.8\% has been demonstrated with randomised bench-marking \cite{Baldwin2019}. With the larger numbers of qubits needed for computation beyond classical capability these gates require longer operation times \cite{Casanova2012}, preventing classically unrealisable computation within the decoherence time. 

\begin{figure}
\includegraphics[width=0.95 \columnwidth]{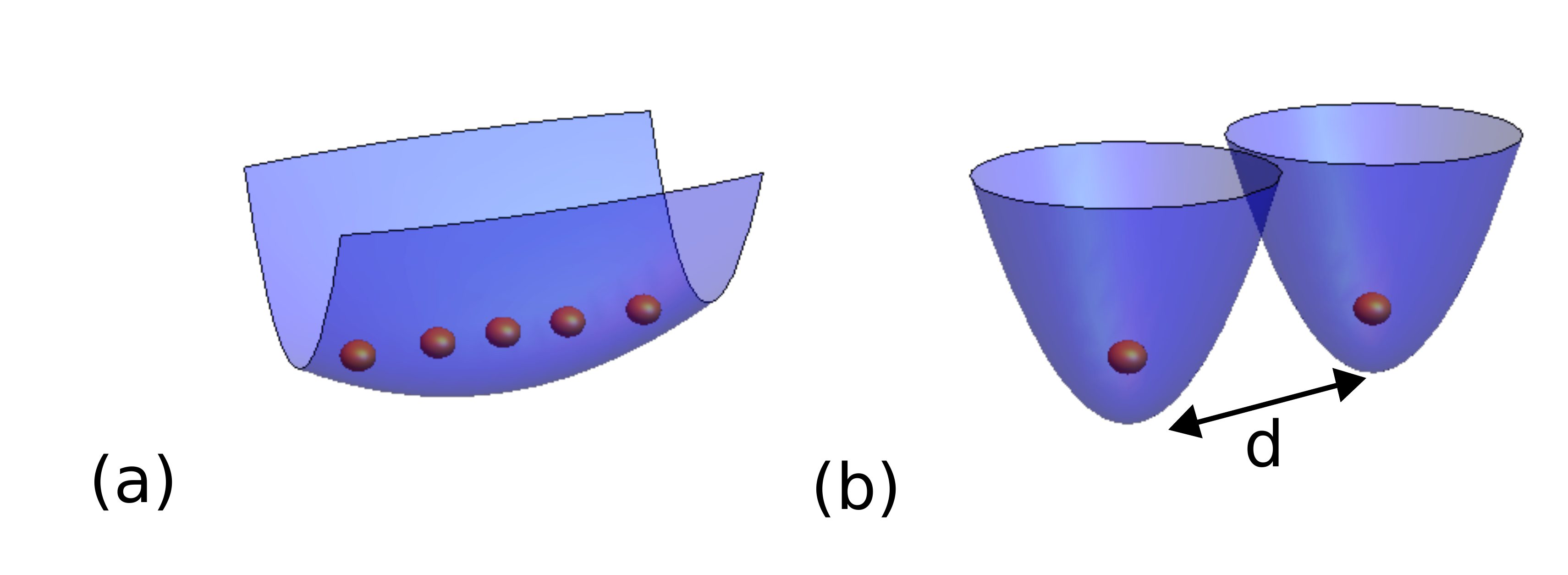}
\caption{(a) Diagram of a linear ion chain, as would be found in linear Paul trap experiments. The lasers used to conduct the two-qubit gates considered in this paper are applied orthogonal to the axis along which the ions are aligned. (b) Diagram of the ions trapped in individual microtraps. In this case, the gates considered are conducted using lasers parallel to the ion chain.}
\label{fig:traps} 
\end{figure}

Fast gates using ultrafast pulses are poised to resolve this situation \cite{Garcia-Ripoll2004,Duan2004,Taylor2017,Bentley2015}. These gate schemes use state-dependent kicks (SDKs) from ultrafast $\pi$-pulses to the ions, inducing state-dependent motion. This results in a state-dependent phase shift mediated by the Coulomb interaction. High fidelity gates 100-1000 times faster than sideband-resolving gates can theoretically be achieved with current laser technology \cite{Taylor2017,Ratcliffe2018}. schemes with ultrafast pulses have recently been used to create Schrodinger cat states \cite{Johnson2017} and to prepare bell states \cite{Wong-Campos2017}. Similar schemes using an amplitude-modulated continuous laser creating a dipole force to prepare bell states have also been demonstrated \cite{Schafer2017}. So far, these schemes have only been applied to linear Paul traps in which the ions are arranged linearly in a common trap. This architecture poses some difficulties in scaling to larger numbers of qubits.  When the number of ions in the trap is increased, the longitudinal trapping frequency must be lowered to prevent buckling of the ion chain, which then slows both adiabatic and fast gates conducted on the longitudinal motional modes \cite{Shimshoni2011,Taylor2017}. 

%Fast gates can operated between individual microtraps maintaining a tight trapping, with a fidelity that asymptotes as the ion number increases without requiring re-optimisation \cite{Ratcliffe2018}. Here, we examine gates between ions in separate microtraps using their longitudinal motion and also between ions in the same linear Paul trap using their radial modes.

Ion traps use an oscillating RF potential to generate a 3D trapping potential. The resulting dynamics approximate a simple harmonic oscillator, but with the addition of a rapid oscillation.  This fast oscillation induced by the oscillating RF potential is referred to as micromotion \cite{Leibfried2003}, or intrinsic micromotion.  While for a perfect trap it is entirely deterministic and entirely in the radial plane, trap imperfections cause this motion to be complicated and coupled into the axial modes.  This is typically called ``excess micromotion'', and considerable effort goes into reducing this in experiments.  Even in perfect traps, the ``intrinsic'' micromotion still exists.  It has largely been treated as an undesirable effect, as it complicates the use of the radial mode, which would otherwise provide the benefit of operating at a higher trapping frequency.  This effect is minimised for adiabatic entangling gates, where it has been shown that it does not inhibit the ability to apply high fidelity entangling operations for sideband-resolving geometric phase gates using a continuously shaped laser pulse \cite{Shen2014a} or by addressing the micromotion sidebands \cite{Gaebler2016,Bermudez2017}.  

We show that fast gates are more affected by micromotion, and that gates designed without awareness of the micromotion have dramatically reduced fidelity.  We speculate that micromotion may have been a contributing factor to the low fidelities observed in \cite{Wong-Campos2017}.  Micromotion would also impact the operation of gates in 2D architectures such as microtraps \cite{Kumph2016}, as the coupled modes in the gate scheme cannot all be orthogonal to the micromotion axes.

Fortunately, designing fast gates with foreknowledge of the micromotion leads to good news: we find that the fidelity can be \textit{increased} for some experimentally accessible parameter regimes.  It is interesting that the state-independent forces in micromotion can in fact enhance the rate of phase accumulation from the state-dependent forces of the laser pulses.  This improvement is present both in the radial modes of a linear Paul trap and in the transverse modes of a microtrap system, and exists for all trap parameters.  We also show that these gates are robust to realistic sources of experimental error.  

\section{\label{sec:model}Model}

Ions can only be trapped in 3D by a time-averaged potential. The time-dependent potential used to generate this trapping is given by:
\begin{eqnarray}
	\Phi(x,y,z,t)&=&\frac{U}{2}(\alpha x^2+\beta y^2+\gamma z^2) \label{eq:potential} \\ 
	   &+&\cos{(\omega_{\text{RF}} t + \phi_{\text{RF}})}\frac{\tilde{U}}{2}(\alpha^\prime x^2+\beta^\prime y^2+\gamma^\prime z^2), \nonumber
\end{eqnarray}
which has previously been shown to lead to motion of the following form \cite{Leibfried2003}:
\begin{eqnarray}
    x(t)&=&A e^{i \frac{\beta \omega_{\text{RF}}}{2} t} \sum_{j=-\infty}^{\infty} C_j e^{i j ( \omega_{\text{RF}} t +  \phi_{\text{RF}})} \nonumber \\ &+& B e^{- i \frac{\beta \omega_{\text{RF}}}{2} t} \sum_{j=-\infty}^{\infty} C_j e^{-i j ( \omega_{\text{RF}} t + \phi_{\text{RF}})},
\end{eqnarray}
where $A$ and $B$ are arbitrary constants determined by boundary conditions, $\beta_x$, $a_x$, and $q_x$ are given by
\begin{eqnarray}
\beta_x \approx \sqrt{a_x + \frac{q_x^2}{2}}, a_x=\frac{4 Z |e| U \alpha}{m \omega_{\text{RF}}^2}, q_x=-\frac{2 Z |e| \tilde{U} \alpha^\prime}{m \omega_{\text{RF}}^2}, \nonumber
\end{eqnarray}
and the coefficients $\lbrace C_j \rbrace$ are determined by a continued fraction with respect to $\beta_x$, $a_x$, and $q_x$  \cite{Leibfried2003}. This equation describes a motion with a secular trapping frequency of $\omega=(1/2) \beta_x \omega_{\text{RF}}$, and an additional high-frequency oscillation which is referred to as micromotion. To ensure sufficient numerical convergence of the optimised gates in this work, it is necessary to take the above expansion over at least $j=-3$ to $j=3$ for values of $q$ greater than $\sim 0.3$. In the work presented here the value $\beta_x$ is found using the MathieuCharacteristicExponent function provided by Mathematica. This gives an accurate value for the exponent for all values of $a$ and $q$.  

In this work we considered fast gates conducted between neighbouring ions in a linear chain of microtraps using the motional modes aligned parallel to the chain of ions, and gates between neighbouring ions in a linear Paul trap using their radial modes. This is shown in Fig.~\ref{fig:traps} (a) \& (b) respectively, and the potentials describing these architectures are given by $V_{M}$ and $V_{P}$ respectively:

\begin{eqnarray}
	V_{M} &=& \frac{e^2}{4 \pi \varepsilon_0} \sum\limits_{i=1}^{N-1} \sum\limits_{j=i+1}^{N} \frac{1}{((j-i)d + x_j - x_i)} \nonumber \\ 
	&+&\frac{1}{2} M \frac{a - 2 q \cos{(\omega_{RF} t)}}{4} \omega_{RF}^2 \sum\limits_{i=1}^N  x_i^2 
    \label{eq:poten_micro}
\end{eqnarray}

\begin{eqnarray}
	V_{P} &=& \frac{e^2}{4 \pi \varepsilon_0} \sum\limits_{i=1}^{N-1} \sum\limits_{j=i+1}^{N} \frac{1}{\sqrt{(x_j - x_i)^2+(z_j - z_i)^2}} \nonumber \\ 
	&+& \frac{1}{2} \frac{M}{4} \omega_{RF}^2 \sum\limits_{i=1}^N (a_z z_i^2 + (a_x - 2 q_x \cos{(\omega_{RF} t)} x_i^2) \nonumber \\ \
    \label{eq:poten_paul}
\end{eqnarray} 
where $x_i$ is the deviation of the $i^\text{th}$ ion in the chain from its equilibrium, $d$ is the separation between each microtrap, $\omega_{\text{RF}}$ is the RF drive frequency used to generate the trapping potential, $M$ is the ion mass and $N$ the number of ions in the chain.  We consider gates between neighbouring ions using the modes that are only coupled to the $x$-axis for the microtrap array model, and the $z$-axis in the linear Paul trap model. We will drop the subscripts for $a$, $q$ and $\beta$ for the remaining discussion.

We use a normal mode expansion to describe the classical coupled motion of the ions. This approximates the motion in terms of $N$ oscillatory modes, each mode described by some frequency of oscillation $\omega_p$ and coupling to the ions $\vec{b}_p$. This is done by linearising the potential around the ions' stationary points, and is valid for sufficiently small displacements of the ions around their stationary points. We use the secular trapping period of the common motional mode $2 \pi/\omega_{\text{COM}}$ as the natural time scale as it simplifies the gate analysis, where $\omega_{\text{COM}}$ can be expressed as $\omega_{\text{COM}}= \frac{1}{2} \beta \omega_{\text{RF}}$. We then use the non-dimensional time $\tau$ given by $\tau = \frac{\beta \omega_{\text{RF}}}{4 \pi} t$. This will then give the motion as:

\begin{eqnarray}
	x_i &=& A_p b_p^i e^{i 2 \pi \frac{\omega_p}{\omega} \tau} \sum_{j=-\infty}^{\infty} C_j e^{\frac{i 4 \pi j (\tau +  \phi_{\text{RF}})}{\beta}} \nonumber \\
	&+& B_p b_p^i e^{-i 2 \pi \frac{\omega_p}{\omega} \tau} \sum_{j=-\infty}^{\infty} C_j e^{-\frac{i 4 \pi j (\tau + \phi_{\text{RF}})}{\beta}}
    \label{eq:mot}
\end{eqnarray}
with the amplitudes of each mode $A_p$, $B_p$ and the phase of the RF-drive $\phi_{\text{RF}}$ relative to the time $\tau=0$, determined by choice of initial conditions. We note that to match the phase evolution of the ODE, the first order solution is not sufficient, and we need to take the Fourier expansion in Eq.~\eqref{eq:mot} to the $3^\text{rd}$ term. While the frequency of the common motional mode will remain the same as the secular frequency of the single ion system $\omega$, the motion of the other secular modes will be different to those derived in the absence of micromotion. To determine the motion of the secular modes, it is necessary to solve the Hill equations resulting from the eigen-decomposition of the coupled ODE's linearised about the equilibrium position. The Hill equations for each of the modes will take the form

\begin{eqnarray}
    -\frac{(2 \pi )^2}{\beta ^2} \sum_{j=0}^\infty h_{j,p} \cos \left(\frac{4 \pi  \tau }{\beta } \right) \Upsilon_p = \frac{d^2 \Upsilon_p}{d \tau^2},
    \label{eq:hill}
\end{eqnarray}
where $\Upsilon_p$ is some coordinate describing the modes displacement, and the coefficients $h_{j,p}$ can be found in terms of the equilibrium position: 

\begin{eqnarray}
	h_{j,p} = \int_0^{\beta/2} \frac{4 \beta \lambda  \cos \left(\frac{4 \pi j \tau }{\beta }\right)}{(\delta + 2 x_{0,1}(\tau))^3} d \tau.
    \label{eq:hill_coefs}
\end{eqnarray}
We truncate Eq.~\eqref{eq:hill} this to the first order Fourier term, which is of the form of a Mathieu equation

\begin{eqnarray}
    -\frac{(2 \pi )^2}{\beta ^2} \left( a_p - 2 q_p \cos \left(\frac{4 \pi  \tau }{\beta } \right) \right) \Upsilon_p = \frac{d^2 \Upsilon_p}{d \tau^2},
    \label{eq:hill2}
\end{eqnarray}
where $a_p$ and $q_p$ are now mode-specific. These can then be used to find $\beta_p$, which can then be used to find the secular mode frequency $\omega_p$ as

\begin{eqnarray}
	\omega_p = \beta_p \omega_{\text{RF}}.
    \label{eq:sec_modes}
\end{eqnarray}

In the case of a linear Paul trap the equilibrium remains constant in time and the linearisation results in a change of the parameter $a_p$ for that mode. This will be given by

\begin{eqnarray}
    a_p = a-\frac{\beta ^2 \kappa ^2}{(2 \pi )^2}.
    \label{eq:trans_a}
\end{eqnarray}

For a microtrap array it is necessary to find the periodic crystal solution. This is the motion that the ions will undertake without any excitation of the secular modes, and may be regarded as a form of ``excess'' micromotion in previous analyses. This motion will be periodic with the RF-drive frequency and will take the form

\begin{eqnarray}
	x_{0,i}(\tau) &=& u_{0,i} + \sum_{j=1}^\infty u_{j,i} \cos{\left( \frac{4 \pi j \tau }{\beta }\right)} \nonumber \\ &+& \sum_{j=1}^\infty w_{j,i} \sin{\left( \frac{4 \pi j \tau }{\beta }\right)}.
    \label{eq:equil_mot}
\end{eqnarray}

In particular, the two-ion case will take the form

\begin{eqnarray}
	x_{0,1}(\tau) = u_{0} + \sum_{j=1}^\infty u_{j} \cos{\left( \frac{4 \pi j \tau }{\beta }\right)} \nonumber \\
	x_{0,2}(\tau) = -u_{0} - \sum_{j=1}^\infty u_{j} \cos{\left( \frac{4 \pi j \tau }{\beta }\right)}
    \label{eq:equil_mot_two}
\end{eqnarray}

The coefficients can be found using a continued series of matrix determinants \cite{Landa2012}. In this paper, we use an iterative method, whereby an initial guess of this motion is made and improved over a number of iterations. This is achieved by using the trial solution as the initial condition for an numerical ODE simulation of the motion for a time close to an integer number of secular trap periods. The average position is then used to adjust the constant component of the periodic crystal solution. The Fourier components then form the coefficients $a_n$. We have found that this method converges to the correct periodic crystal solution.

\section{Gate Scheme and Optimisation}

  A description of many of the features of this scheme and optimisation have been previously presented in \cite{Ratcliffe2018} and are presented here for convenience to the reader. Fast gates using ultrafast pulses are implemented by illuminating a pair of ions with a series of counter-propagating $\pi$-pulse pairs. Each of these counter-propagating pulse pairs induces a state-dependent momentum kick on the incident ions, appearing as vertical jumps in a phase space diagram. By appropriately choosing the arrival times and arrival orderings of the pulse pairs it is possible to implement an entangling gate. The fully entangling gate operation that can be implemented using this method is the controlled phase (CPhase) gate, also referred to as a controlled-zz gate, given as:
  
  \begin{eqnarray}
	\hat{U}_\text{CPhase} = e^{i \frac{\pi}{4} \sigma^z_1 \sigma^z_2}.
    \label{eq:cphase}
\end{eqnarray}
  
To enable a more tractable optimisation of the gate fidelity, the pulse pairs are organised into groups of pulses. These groups are assumed to provide an instantaneous momentum consisting of $\vec{z}$ pulse pairs at times $\vec{t}$. We use a gate scheme that is a generalisation of the Fast Robust Anti-symmetric Gate (FRAG) scheme \cite{Bentley2015}, which is in turn a variant of the GZC scheme \cite{Garcia-Ripoll2003}. It consists of six groups of counter-propagating $\pi$-pulses incident on the ions to be entangled. The timings of theses pulses and the number of pulses in each pulse group are given by the vectors $\underline{t}$ and $\underline{z}$ respectively:

\begin{eqnarray}
	\underline{t} &=& (-\tau_1, -\tau_2, -\tau_3, \tau_3, \tau_2, \tau_1), \nonumber \\ 
    \underline{z} &=& (-n, 2n, -2n, 2n, -2n, n). 
    \label{FRAG}
\end{eqnarray}

The sign of the components of $\underline{z}$ corresponds to changing the direction of the initially incident pulse, and the factor of $n$ is an integer that characterises the overall scale of numbers of pulses in each pulse group. 

To produce an high-fidelity CPhase gate the timings ($\tau_1,\tau_2,\tau_3$) are optimised to give the desired gate.  In the original FRAG scheme proposal there was a strict ordering on the magnitude of ($\tau_1,\tau_2,\tau_3$). In this implementation we do not impose a strict ordering of the times ($\tau_1,\tau_2,\tau_3$), effectively resulting in a set of six possible pulse schemes.  The total gate time $\tau_G$ is therefore twice the maximum of the values of $\tau_1$, $\tau_2$, and $\tau_3$.

The optimisation was carried out by using a large number of local optimisations within a bounded gate time. The bounds are then incrementally increased and the new region is searched. Using this method, we establish a relationship between the allowed gate time and the gate infidelity. 

\begin{figure}
\label{fig:FG} 
\includegraphics[width=1 \columnwidth]{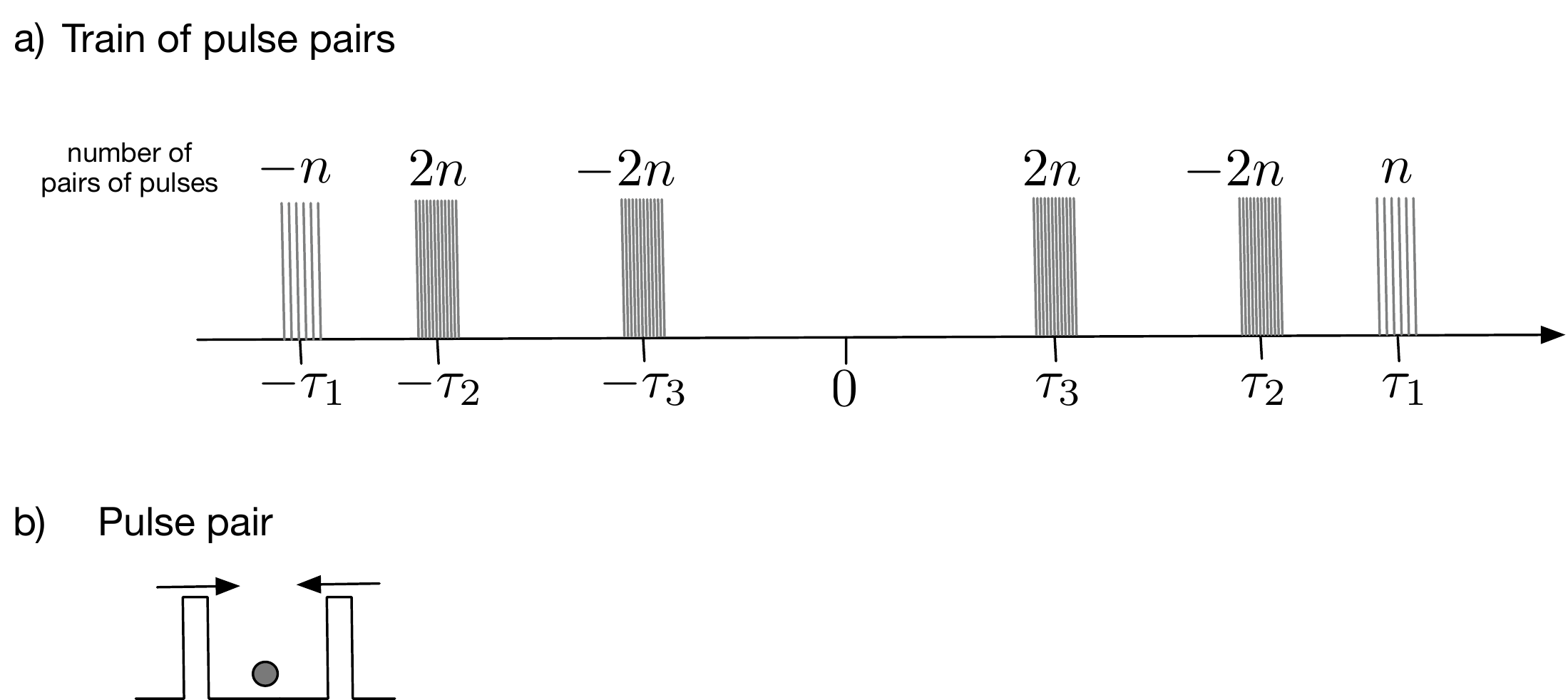}
\caption{a) Diagram with the pulse timing for the FRAG scheme. The components $z_j$ of the $z$ vector indicate the number of pairs of pulses that hit the ion at each time $\tau_j$. The sign in $z_j$ indicates which pulse within each pair (shown in b) reaches the ion first. This gives the sign of the momentum kick imprinted on the ion.}
\end{figure}

We use the state-averaged fidelity $F$ as the measure of a gate's performance in this work because it can be calculated with high efficiency \cite{Bentley2015,Bentley2016,Ratcliffe2018}. Inclusion of the micromotion adds additional terms to the analytic fidelity expression, as detailed in Appendix~\ref{sec:mic}. As we are interested in gates with fidelities close to unity, it is sensible to report this in terms of the infidelity $1-F$. We then simplify this expression for small errors in phase and motional restoration, giving the infidelity expression as
\begin{eqnarray}
	1 - \text{F} &\approx& \frac{2}{3} \Delta\phi^2 \nonumber \\
	&+& \frac{4}{3} \sum_p \left(\frac{1}{2} + \overline{n}_p\right) \left((b_p^1)^2 + (b_p^2)^2\right) \Delta {P_p}^2,
    \label{costFun}
\end{eqnarray}
where $\Delta \phi$ is the error in the phase, $\Delta P_p$ is the displacement of the $p^\text{th}$ mode in phase-space, $b_p^n$ is the coupling of the $p^\text{th}$ mode to the $j^\text{th}$ ion, and $\overline{n}_p$ is the mean phonon occupation of the $p^\text{th}$ mode. Expressions for $\Delta \phi$ and $\Delta P_p$ will be given later in Eq.~\eqref{eq:gate_errors}. We assume a thermal product state, with a mean mode occupation of $\overline{n}_p=0.1$ throughout this work. This assumption has been the basis throughout previous works \cite{Ratcliffe2018,Bentley2015} and has been experimentally demonstrated \cite{Inlek2017}. For larger mode occupations the infidelity will grow linearly with $\overline{n}_p$. The final quoted infidelities were confirmed by directly integrating the ODEs for the classical phase space trajectories over the qubit basis states. The enclosed signed area of these trajectories is used to calculate the geometric phase, also known as the Berry phase.

For the purposes of designing fast gates for two-ion systems, traps are well characterised by the dimensionless parameter $\chi$, which is the scaled difference between the breathing and common motional modes $\chi=\frac{\omega_{\text{BR}}-\omega}{\omega}$ in the direction of the laser-induced motion \cite{Ratcliffe2018}.  Expressions for $\chi$ in terms of trap parameters depend on the geometry, and are given in Appendix~\ref{sec:param}. In a linear Paul trap $\chi$ is negative for motion in the radial modes, indicating a phase acquisition rate of the opposite sign to gates conducted using the axial modes.  In this case, we optimise for a controlled phase gate with opposite relative phase, which is equivalent up to single qubit $\pi$ rotations.  

\begin{figure*}
	\subfloat[Without Micromotion]{ \includegraphics[width=0.45\linewidth]{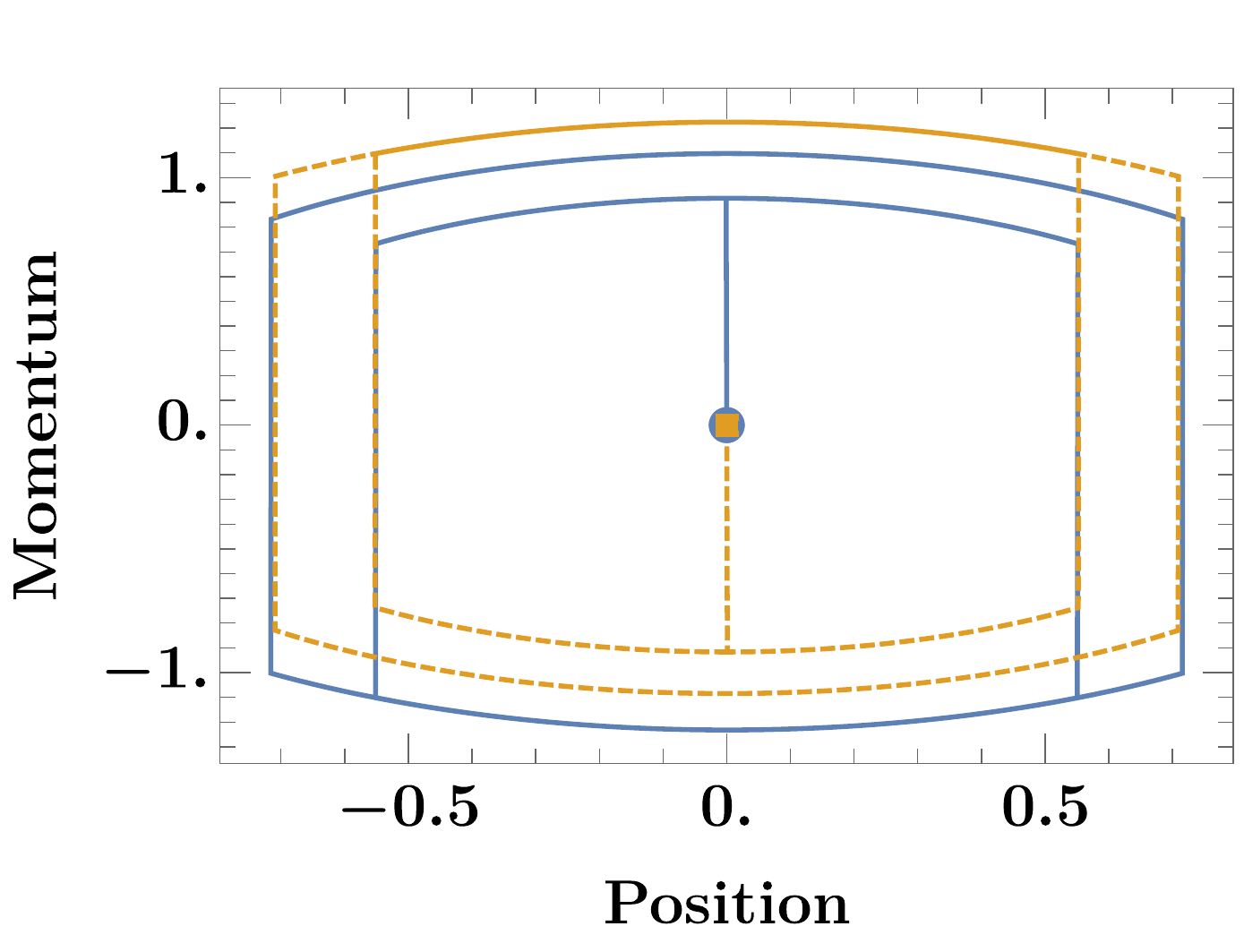} } 
	\subfloat[With Micromotion]{ \includegraphics[width=0.45\linewidth]{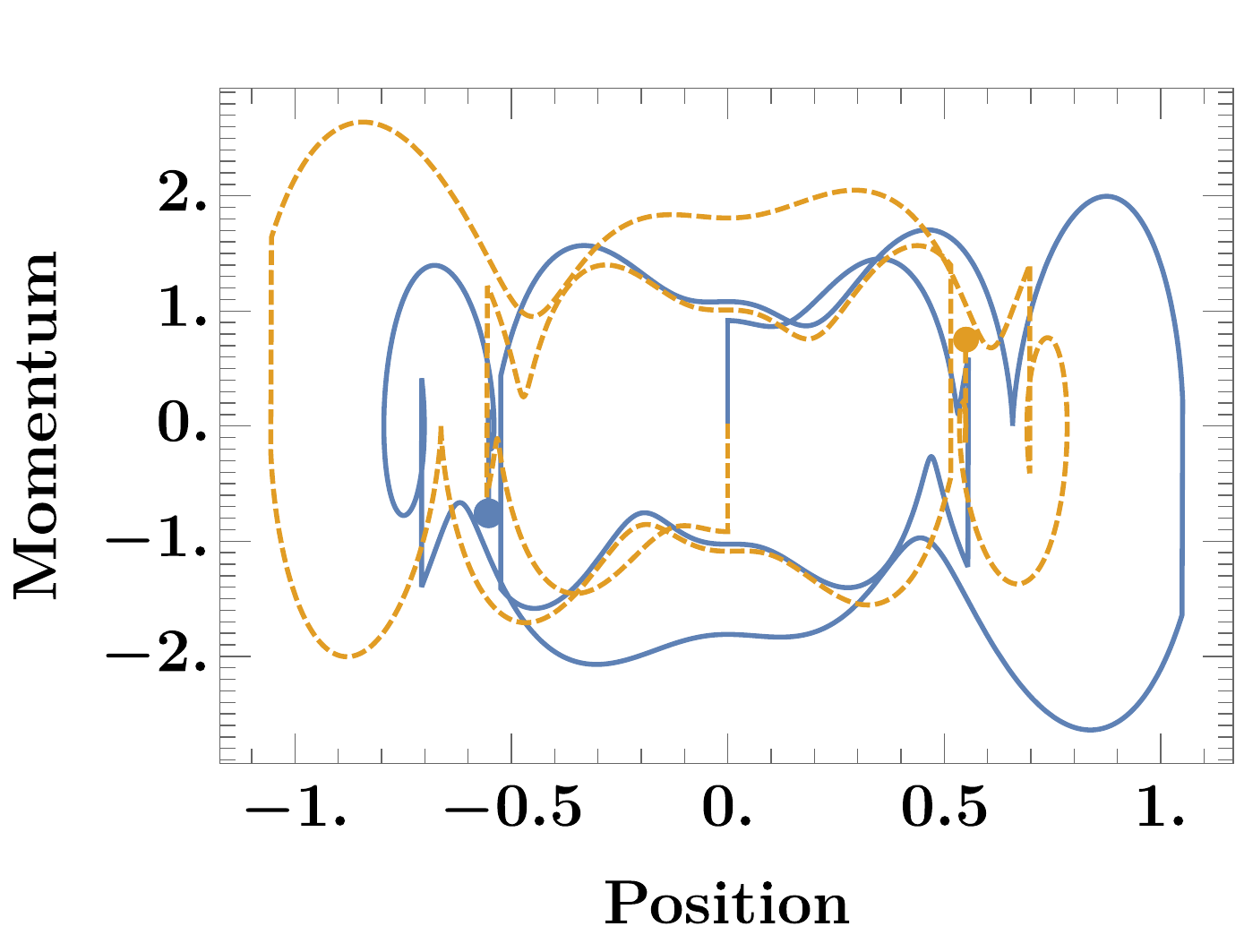} } 
\caption{\label{phase_plot} Phase space trajectories of the common motional (blue solid) and rocking (orange dashed) modes of a two ion fast gate conducted using the radial modes of a linear Paul trap. (a) Trajectories without micromotion, showing a clear motional restoration indicated by the circular markers at the origin of the plot. The theoretical infidelity of this gate is roughly $10^{-12}$. (b) The same fast gate operation with the inclusion of micromotion. This clearly shows that there is no longer restoration of the modes, indicated by the two markers showing the end of the trajectory far from the origin. The parameters $a$ and $q$ were set to $0.0$ and $0.2$ respectively, the theoretical infidelity of this gate is now $\sim 0.5$.}.
\end{figure*}

The phase space trajectories are more complicated for the system evolving with micromotion, this comparison is shown in Fig.~\ref{phase_plot}. These trajectories can be averaged to remove the rapid oscillation induced by micromotion. In the case where a continuous phase space displacement is applied to the system, the averaged trajectory will be well approximated by the trajectory of the approximate harmonic trap, provided the interaction is slow with respect to the RF drive. However, when these interactions occur on timescales faster than the RF drive, the averaged phase-space trajectories can differ substantially from those that would be found using a simple harmonic trapping approximation. The SDKs used in this work provide a near instantaneous momentum kick, considerably faster than the RF oscillation. The averaged phase space displacements for these kicks will be larger or smaller depending on when they are timed in the RF cycle. We define $\mu$ as the relative increase in the maximum displacement of the ions $D$ over the maximum displacement of the ions in a simple harmonic potential $D^\prime$, achieved through a momentum kick. This will be given as:

\begin{eqnarray} 
    \mu &\approx& \left(1-\frac{2 \left(\beta ^2+4\right) q \cos (\phi_{\text{RF}} )}{\left(\beta ^2-4\right)^2}\right)^2. 
\end{eqnarray}

By timing the arrival of the SDKs to particular times in the RF cycle, it is possible to achieve faster gate times using the same number of counter-propagating pulse pairs. Here we investigate a scheme in which the gate is optimised assuming all pulses arrive at the same point in the RF cycle. This assumption simplifies the infidelity expression with the inclusion of micromotion:

\begin{eqnarray} 
    \Delta P_p &=&2 \mu \sqrt{\frac{\omega}{\omega_p}} \sum_{k}{z_k \sin{  (\omega_p t_k)}},  \\
    \Delta\phi &=& \left\vert \sum_p{8 \eta^2 \mu \frac{\omega}{\omega_p} b_p^1 b_p^2 \sum_{i\neq j}{z_i z_j \sin{(\omega_p \lvert t_i - t_j \rvert )}}} \right\vert - \frac{\pi}{4}. \nonumber
\label{eq:gate_errors}
\end{eqnarray}

The factor $\mu$ in the acquired phase corresponds to the change in the extent of the phase space displacements due to locking the $\pi$-pulses to a particular point in the RF cycle. The presence of this factor in the motional restoration terms would decrease the fidelity of gates optimised without this factor. When it is included in the optimisation, the ideal achievable fidelities are similar to the system in the absence of micromotion. It does however increase the sensitivity of the scheme to systematic pulse timing errors.  This is an inherent feature of achieving faster gate times, which necessitate larger displacements in phase space.

\begin{figure}
	\subfloat{ \includegraphics[width=0.7\linewidth]{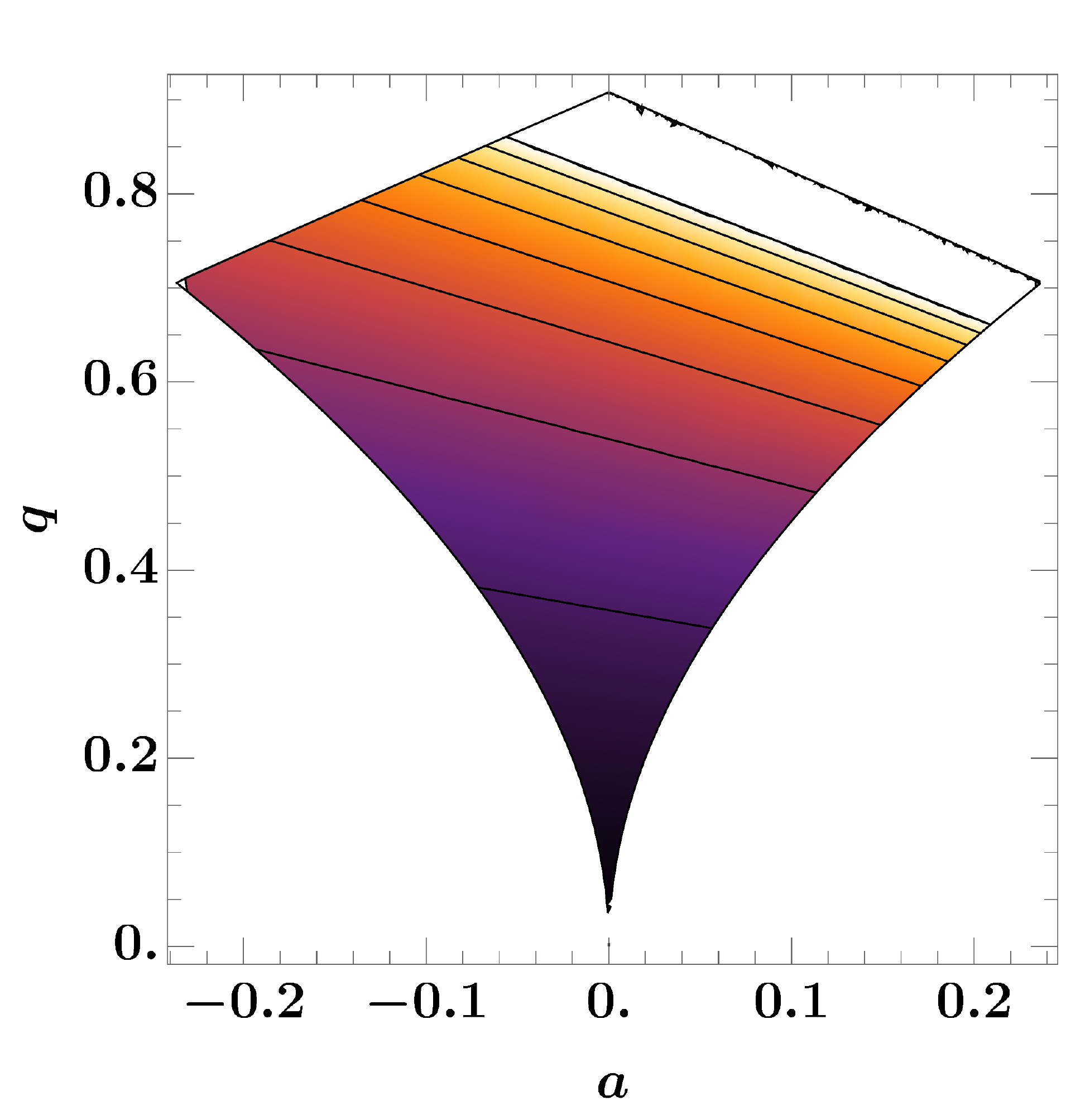} } 
	\subfloat{ \includegraphics[width=0.12\linewidth]{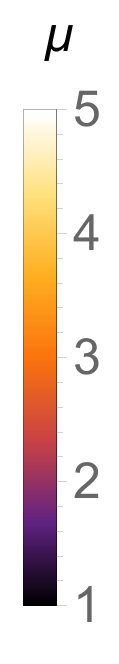} } 
\caption{\label{param_space} The values of $\mu$ are shown as colour gradients over the stable region of trapping parameters $a$ and $q$. The region to the top right of the image diverges towards infinity, but this is generally not a useful region as it results from RF frequencies close to the trapping frequencies.  }
\end{figure}

The values of the parameter $\mu$ are shown in Fig.~\ref{param_space} for values of $a$ and $q$, with the perimeter of the plot marking out the stable trapping regions of the ideal trap. It is worth noting that not all combinations of $a$ and $q$ presented in Fig.~\ref{param_space} are stable in practice. This is due to the presence of resonances induced by the non-linearity of the trapping potential or trap imperfections \cite{Drakoudis2006,Alheit1996,Collings2000}. This typically limits experiments to parameter choices with a value of $\mu$ less than 2.5. The full expression of $\mu$, and its derivation are given in Appendix~\ref{sec:mic}.

\section{Impact of Micromotion and Micromotion Enhancement}

\begin{figure*}
	\subfloat[No DC Offset]{ \includegraphics[width=0.41\linewidth]{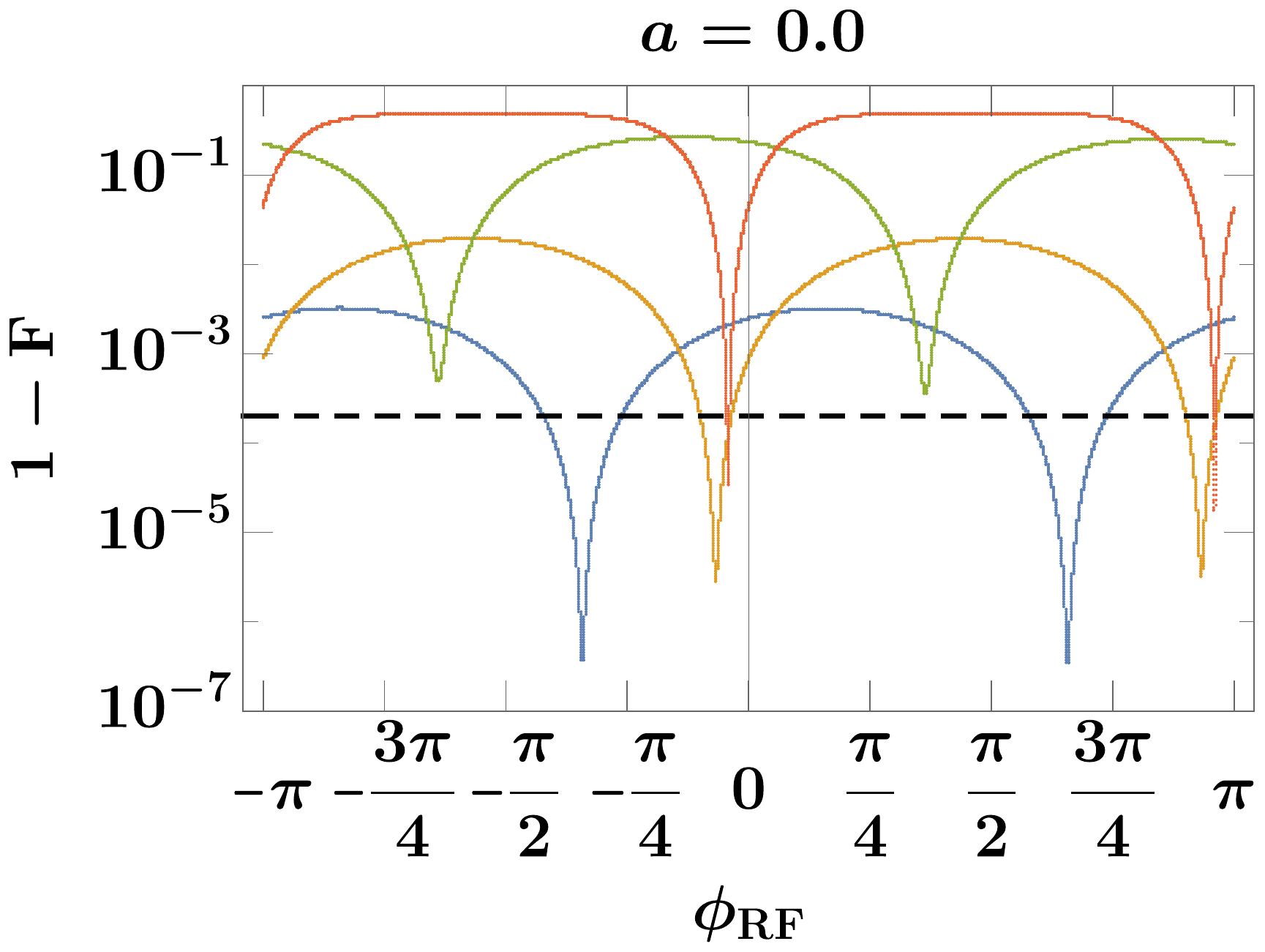} } 
	\subfloat[DC Offset For Fixed Frequency]{ \includegraphics[width=0.41\linewidth]{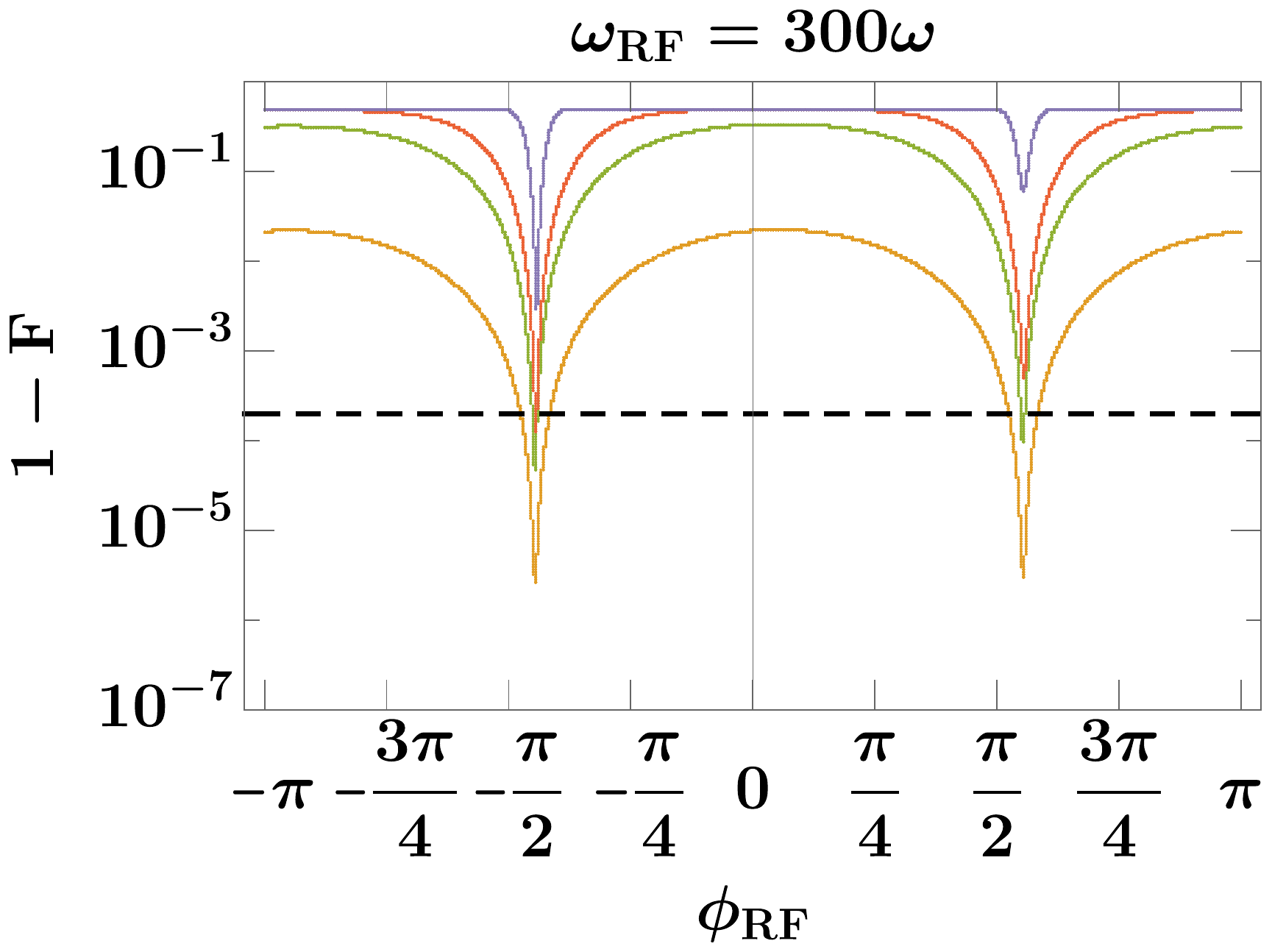} } 
	\subfloat{ \includegraphics[width=0.13\linewidth]{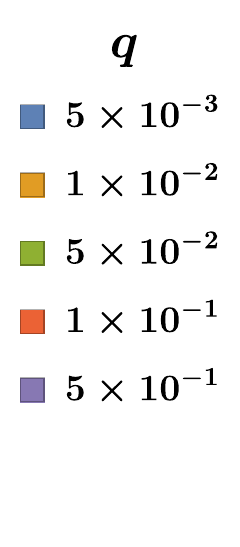} } 
\caption{\label{microkills} Gates that are optimised assuming an absence of micromotion typically have a decreasing fidelity under realistic amounts of micromotion.  We show infidelities for such gates plotted with respect to the phase offset between the first pulse in the gate and the RF drive.  Infidelities are shown for a various values of $q$ for a fixed $q$ and an $a$ chosen to maintain a RF frequency 300 times the secular trapping frequency, using trap parameters $\chi=1.8\times10^{-4}$, $d=100~\mu$m, and $\omega=2\pi$~MHz. Ideal gate fidelities without micromotion were roughly $10^{-12}$, not shown on the scale of these plots.  Increasing values of $q$ indicate an increasing dynamic trapping potential $\tilde{U}$ and show reduced fidelity. The black dashed line shows the fault-tolerant threshold of $2\times10^{-4}$. The dip in these plots corresponds to an optimum in minimising the effects of micromotion. This optimum remains constant when the pulses occur at fixed points within the RF cycle, which is evidenced by the concurrence of these optima when the RF frequency is held constant with respect to the trapping frequency, as in (b). }
\end{figure*}

We now look at the effect micromotion has on a two-qubit gate that was optimised for a system without micromotion. We observe a decreasing fidelity as the dynamic trapping potential $\tilde{U}$ is increased, seen in Fig.~\ref{microkills}.  The dips in these plots corresponds phases that minimise the effects of micromotion on motional restoration, producing closed phase space trajectories. Taking into account the effects of finite repetition rate would further reduce the fidelities shown in Fig.~\ref{microkills}. As will be shown later in this manuscript, micromotion makes fast gate schemes particularly sensitive to the finite repetition rate of the SDKs. A successful implementation of a fast gate using state-dependent kicks therefore requires both RF-induced micromotion to be accounted for in the optimisation process, and either a finite repetition rate sufficiently faster than the RF frequency, or taking the finite repetition rate directly into account in the optimisation process. The latter is typically computationally prohibitive.

\begin{figure*}
	\subfloat[Microtrap (Axial Modes)]{ \includegraphics[width=0.44\linewidth]{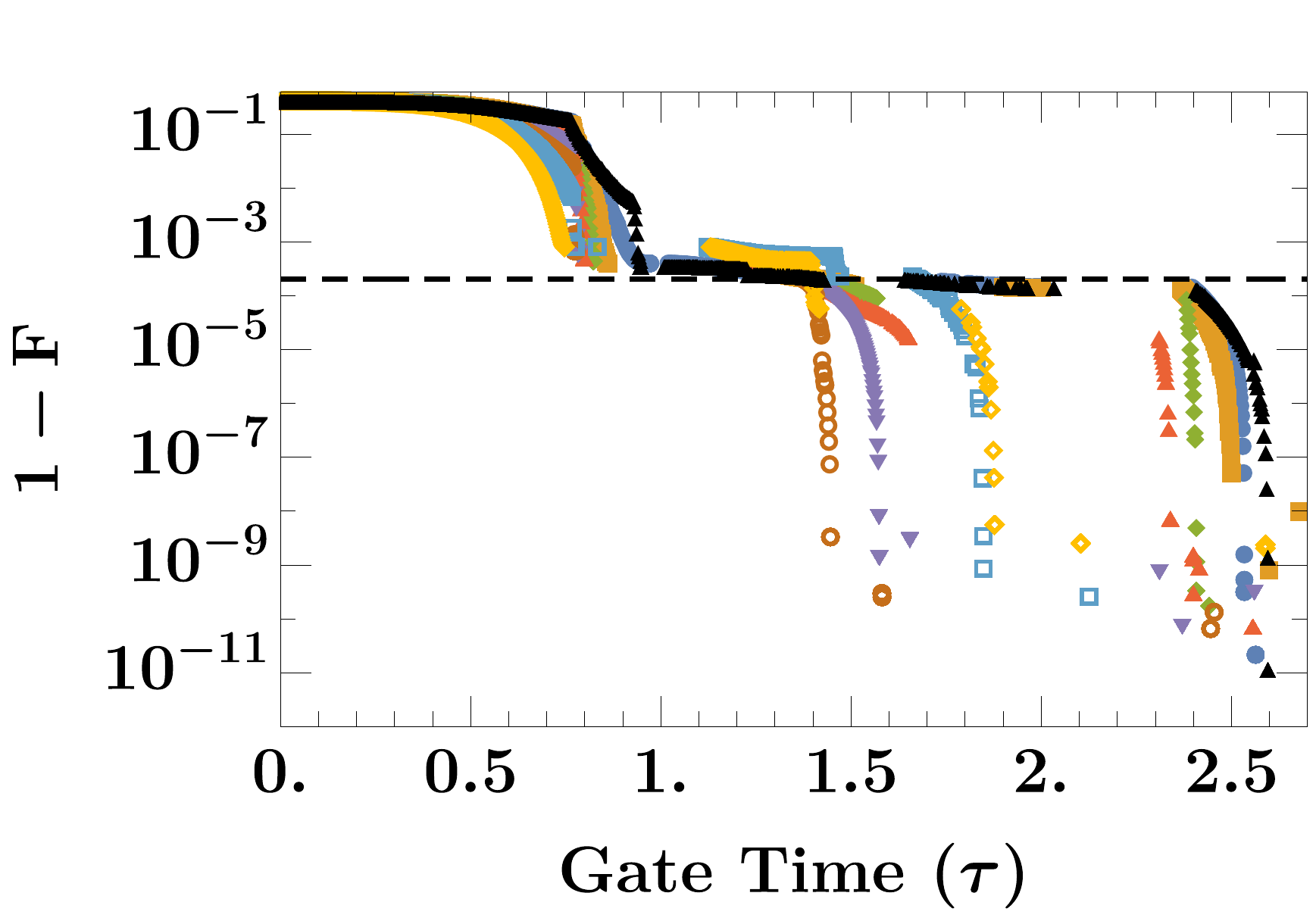} } 
	\subfloat[Linear Paul Trap (Radial Modes)]{ \includegraphics[width=0.44\linewidth]{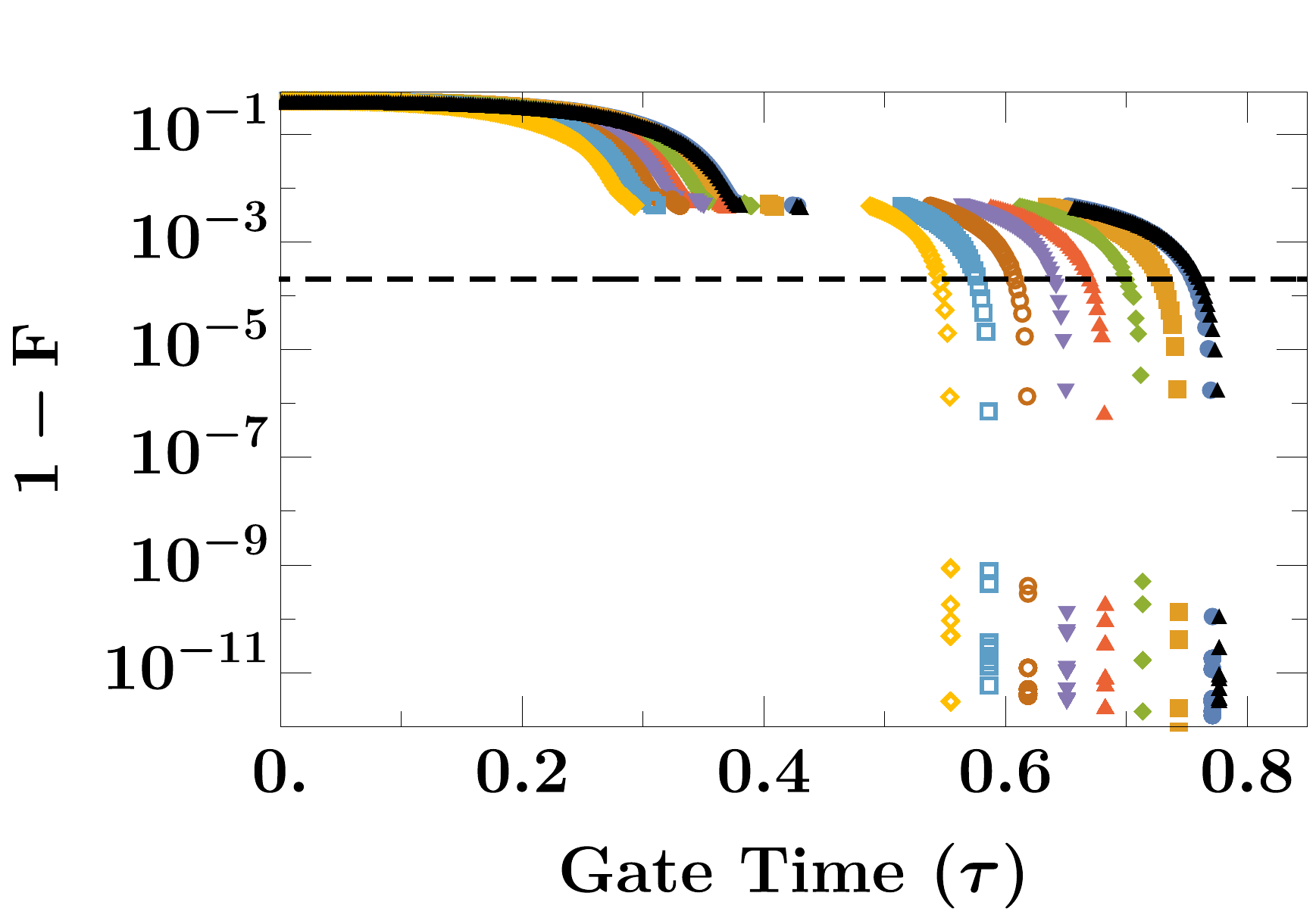} } 
	\subfloat{ \includegraphics[width=0.07\linewidth]{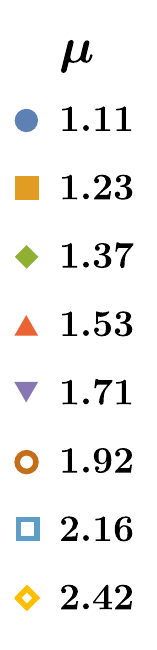} } 
\caption{\label{microImprove} Infidelities of optimised two-qubit gates with pulses occurring at $\pi$ phase to the RF drive, equivalent to a repetition rate locked to the trap RF drive frequency at this phase. Shown with increasing dynamic trapping potential $\tilde{U}$, indicated by increasing values of $\mu$. (a) Microtrap architecture with $\chi=1.8\times10^{-4}$ ($d=100~\mu$m, $\omega=2\pi$~MHz) (b) Linear Paul trap using radial modes with $\chi=-1.4\times10^{-2}$ ($\omega_{A}/\omega_{R}=1/6$). The results shown as black triangles indicate the infidelities for equivalent systems without the inclusion of micromotion. This shows that both systems benefit from a clear micromotion enhancement. The black dashed line shows the fault-tolerant threshold of $2\times10^{-4}$. The plateaus in the infidelity observed for some gate times can be attributed to the anti-symmetric nature of the gate scheme, which is not favoured for gate times that are multiples of half a trapping period. While there is not a monotonic progression in (a) for some values of $\mu$, this is due to the proximity of these results to a half multiples of the trapping period. These trends show that fast gates conducted with the aid of micromotion can be implemented with shorter gate times than gates conducted in the absence of micromotion. }
\end{figure*}

We now examine gates that are designed to operate in the presence of micromotion. The rate of phase acquisition will be maximised when the pulses occur at the point in the RF cycle where $\mu$ is maximised, corresponding to the time at which the oscillating part of the potential is maximally trapping. We thus examine the case where the repetition rate is locked to the RF-drive with a phase between them of $\phi_{\text{RF}}=\pi$.   The phase $\phi_{RF}$ is defined in Eq.~\ref{eq:potential} as the phase of the driving potential at $t=0$, and this assumption means that it will be the phase of the driving potential for all of our pulses.

We find that the performance of two-qubit gates improves with the amount of micromotion, indicated by the infidelity decreasing with increasing $\mu$, which corresponds to an increasing $a$ and $q$.  This improvement is shown in Fig.~\ref{microImprove} for both a microtrap array, and the radial modes of a linear Paul trap. Although only one choice of gate parameter $n$ is shown, the improvement is present for all choices of $n$. The plateaus in the infidelity observed for some gate times can be attributed to the anti-symmetric nature of the gate scheme, which is not favoured for gate times that are multiples of half a trapping period \cite{Ratcliffe2018}. The improvement is more significant for higher values of $q$ although it is not monotonic in the proximity of gate times that are multiples of half trap periods. 

As stated at the end of Section III, not all choices of $a$ and $q$ depicted as stable in Fig.~\ref{param_space} will be stable in experiments. Stable trapping can be achieved with values of $q \sim 0.5$ which are used in current microtrap array implementations \cite{Kumph2016}. However, for these values to be useful for this scheme, a very careful selection of $a$ and $q$ is required in this region to appropriately align troughs in the RF oscillations to the required pulse timings. This degree of control and freedom may be difficult to achieve in practice, which may limit the potential gate improvements.  Even in those cases, it is still important to design the gates to account for the micromotion.

\section{Finite Repetition Rate}

\begin{figure}
\includegraphics[width=0.9 \columnwidth]{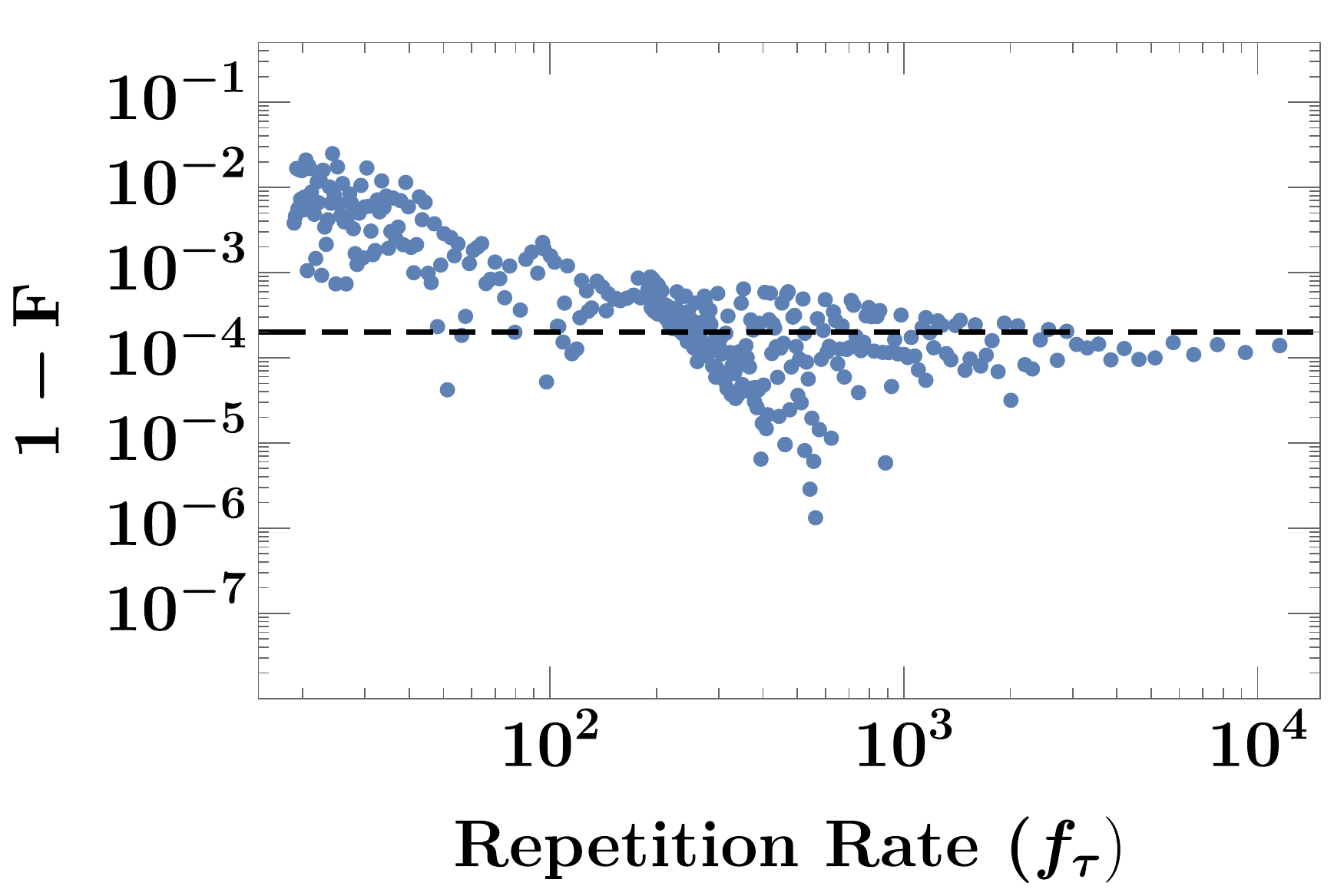}
\caption{ The fidelity of a realistic entangling gate using 50 counter-propagating pulse pairs ($n=5$) and a total gate time of this gate $\tau_G \sim 2~$ trap periods, without a repetition rate locked to the RF-drive. The dynamic trapping parameter $q=0.2$, and the ratio between the RF driving frequency and the secular trapping frequency $f_{\text{RF}} / f_{\tau} \approx 12$. This choice of parameters corresponds to a realistic trapping regime, and shows that our models show robust, high-fidelity gate solutions using current experimental traps.  }
\label{fig:realistic_reps} 
\end{figure}

The optimised gates shown in Fig.~\ref{microImprove} assume perfect phase locking between the pulses and the RF drive, as well as large instantaneous momentum kicks.  In practice, the pulse groups are composed of many small kicks. These small kicks can be made to occur near simultaneously, using a series of optical delay loops \cite{Bentley2012a}. However, this is experimentally complex and requires a high degree of accuracy on the pulse timings \cite{Ratcliffe2018}. A simpler implementation method is to produce pulses at a fixed repetition rate and use a pulse picker to select a subset for use in the gates. We must examine the effects of this approximate implementation to ensure that it does not overly compromise the achievable fidelity of the gate.  

%Restricting the repetition rates to be equal to the RF-drive will, in general, limit the allowable repetition rate dramatically, as RF frequencies are generally not above $100$~MHz, and repetition rates of $5$~GHz \cite{Heinrich2019} have been demonstrated. Thus this requirement would become a limiting factor in gate time. 

Although our gate schemes are designed assuming that all the SDK's in the pulse groups occur instantaneously, in Fig.~\ref{fig:realistic_reps} we examine the effect of implementing them using non-simultaneous pulses achievable with a finite repetition rate.  In order to use physically reasonable choices of RF drive frequency, we adjust the pulse group timings of our gates to ensure our phase-locking condition.  The resulting infidelity assuming an infinite repetition rate was $\sim 1 \times 10^{-4}$. This infidelity could be lowered by optimising over the RF driving frequency. Implementing these gates with a finite repetition rate changes the fidelity significantly, but there are clearly a large number of high-fidelity solutions possible for a realistic trapping parameters, and for $\pi$-pulse repetition rates that have already been demonstrated \cite{Heinrich2019}. 

Perhaps surprisingly, some of the gates implemented with finite repetition rate, which are no longer technically phase locked, were actually better than the phase-locked gates.  The large amount of scatter shows that these gates solutions are highly sensitive to variations in frequency, but even the upper edge of those solutions have promising fidelity.  

\section{Robustness}

\begin{figure}
\includegraphics[width=0.9 \columnwidth]{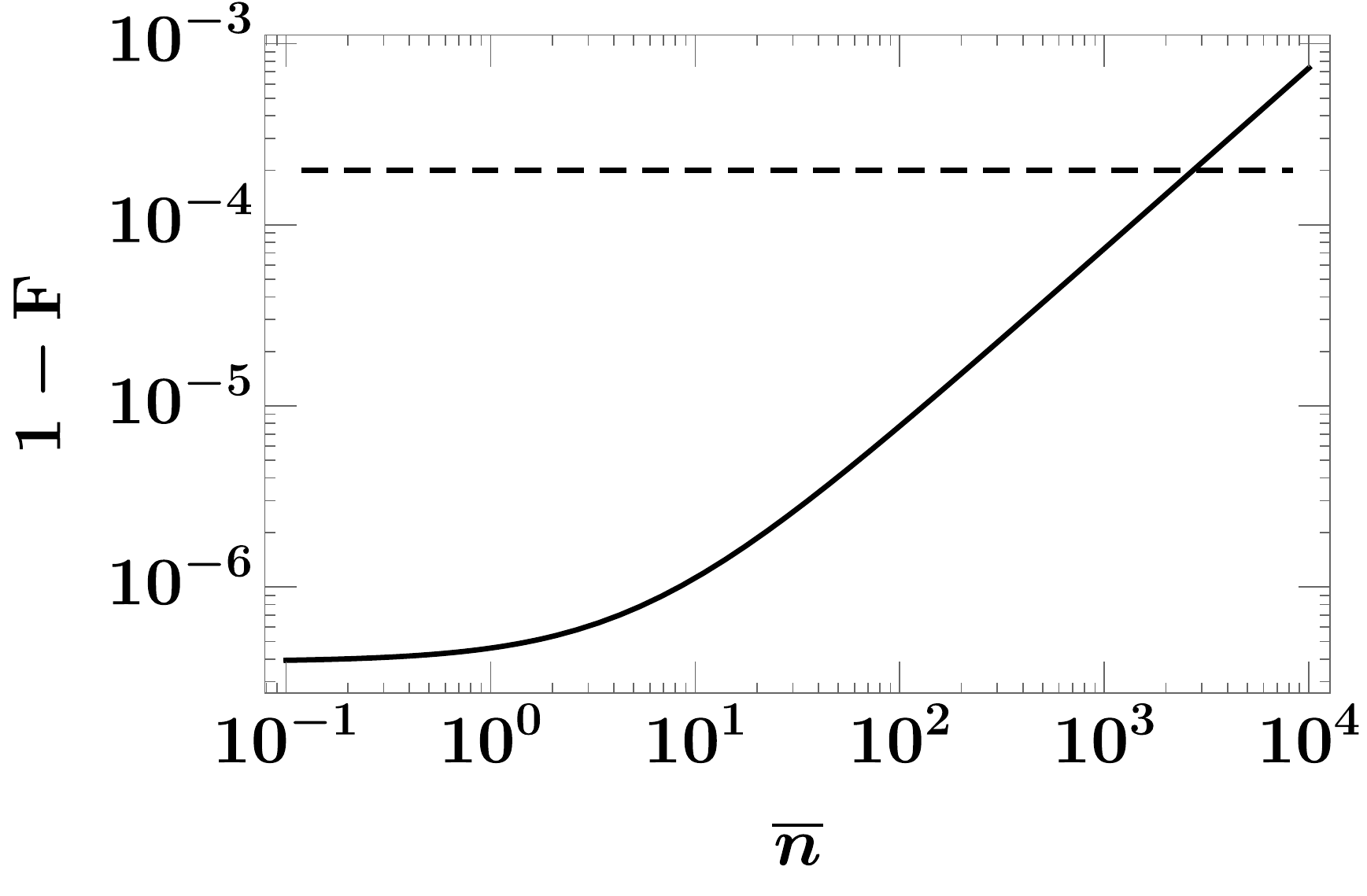}
\caption{ Infidelity of FRAG gates under increasing mean motional occupation $\overline{n}$, assuming the mean occupation is equal across all modes. This shows that these gates, and in-fact any high fidelity gate using ultrafast $\pi$-pulses, remains robust to the temperature of the ion crystal. Certainly this will not be a limiting factor to gate fidelities in the Doppler cooling limit, where $\overline{n} \approx 10$. }
\label{fig:thermal} 
\end{figure}

We now explore the robustness of these schemes to the inevitable imperfections in experimental implementations. 

\textit{Finite temperature:} We begin by first examining the impacts to gate fidelity as a result of higher temperature ion crystals, in contrast ion crystals with a mean motional occupation of $\overline{n}=0.1$ we have assumed until now. If we assume that each mode has the same mean motional occupation, the infidelity of an optimised gate can be expressed as a function of this mean occupation number. This is shown Fig.~\ref{fig:thermal}, where it can be seen that the gate remains robust even with large motional occupation numbers of $\overline{n}=100$. While this is shown for a FRAG gate, this is not unique to the FRAG scheme. This feature will be present in any scheme utilising ultrafast counter-propagating $\pi$-pulses. 

In a similar vein, another form of error can arise due to motional heating during the gate operation. This form of error is investigated in more detail in \cite{Taylor2017}. This work showed that heating events during the gate operation would effectively destroy the gate operation, resulting in very low fidelities. This makes the motional heating an important experimental consideration. Specifically, the motional heating on the modes that are coupled to for a gate operation. As such, for gates operated on the radial modes, the motional heating in the axial modes will have little impact to the fidelity of these gates. And while the motional heating rate generally increases in longer ion chains, this does not necessarily result in an increase in the modes that are coupled during the gate operation.

\begin{figure*}
	\subfloat[Phase Error]{ \includegraphics[width=0.44\linewidth]{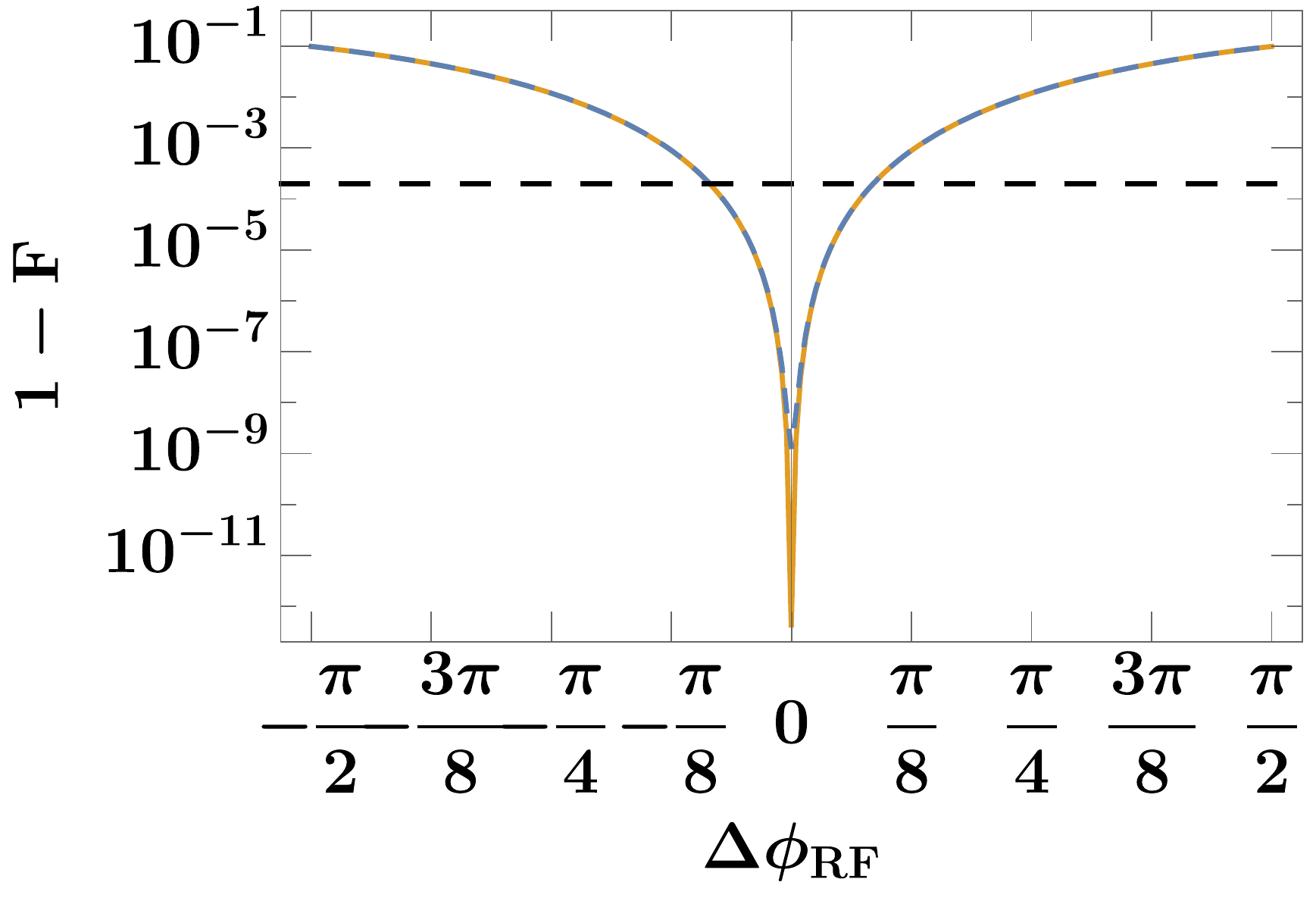} } 
	\subfloat[Trap Parameter Error]{ \includegraphics[width=0.48\linewidth]{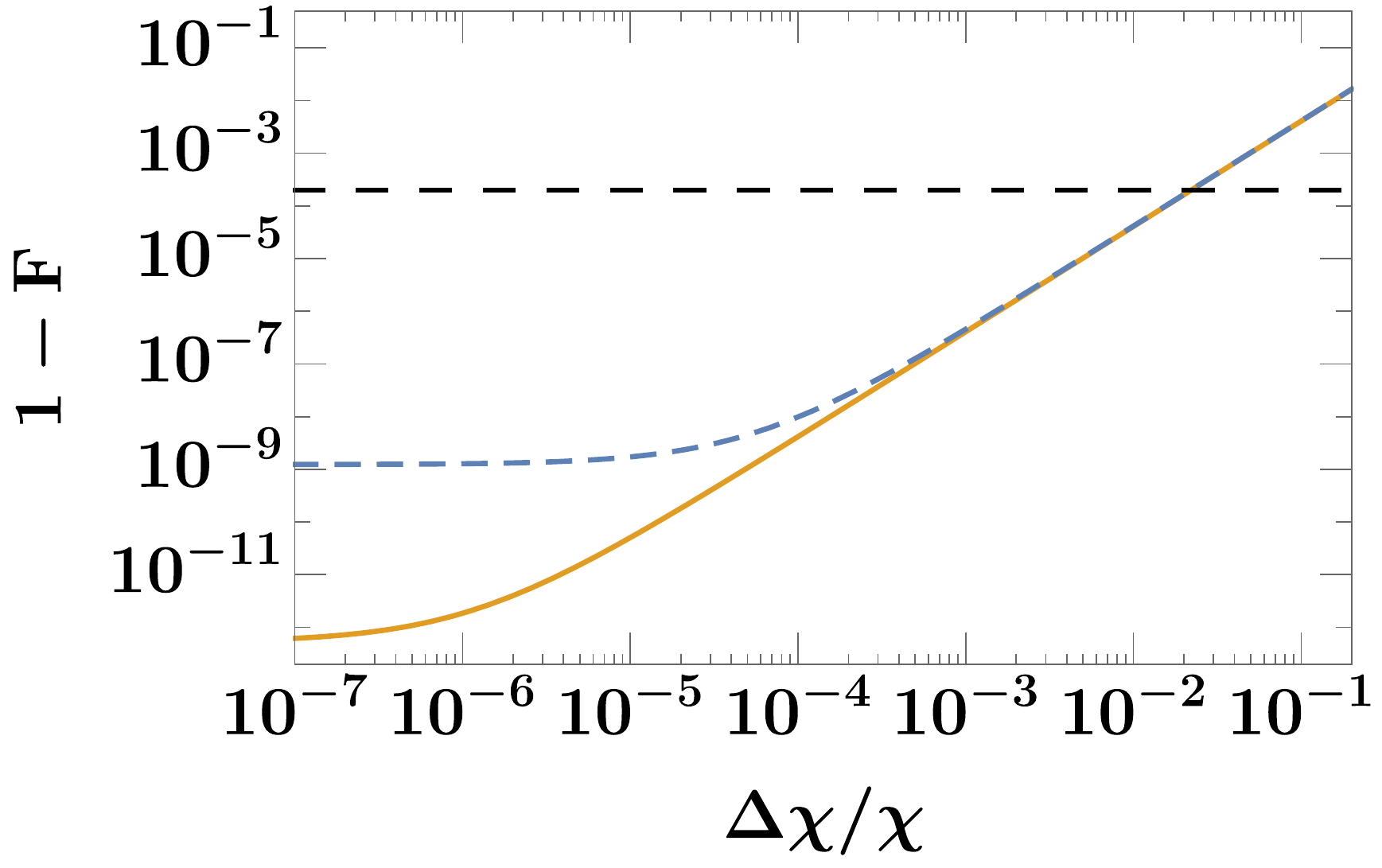} } 
\caption{\label{errors} Plots showing the impact to infidelity arising from realistic experimental errors for a gate with an ideal infidelity of $10^{-9}$ in a microtrap architecture, shown in dashed blue ($\chi=1.3\times1.8\times10^{-4}$, $\tau_G=1.0$, $n=30$, $\mu=2.31$) and a radial mode gate with an ideal infidelity of $10^{-13}$ on a linear Paul trap, shown in orange ($\chi=-1.3\times1.4\times10^{-2}$, $\tau_G=0.7$, $n=12$, $\mu=2.31$). Here $\tau_G$ is the gate time and the multiplying factor of $1.3$ is due to the effects of micromotion. (a) Errors in the phase between laser pulses and the RF-drive. (b) Response to an error $\Delta \chi$ between the real value of $\chi$ and the value used to find an optimised gate. The black dashed line shows the fault-tolerant threshold of $2\times10^{-4}$}.
\end{figure*}

\textit{Imperfect phase matching:}  We now examine what happens when the phase between the pulses arriving and the RF-drive is not exactly $\pi$. As seen in Fig.~\ref{errors} (a), gates remain below error correction infidelity thresholds until there are phase mismatches on the order of $1/16$ of an RF period. The largest RF drives encountered in most experiments are on the order of $100$~MHz, equating to a requirement for nanosecond accuracy of the pulse arrival time with respect to the maximum in the RF trapping potential.

\textit{Trap characterisation:}  The next form of experimental error we examine is the effect of imprecise characterisation of the trap, such as the trapping frequencies or distances between microtraps.  Such errors would lead to an incorrect estimate of $\chi$. Fig.~\ref{errors} (b) shows that gates are also robust to such errors, remaining below error correction infidelity thresholds until errors of approximately $2\%$ of the true value of $\chi$, corresponding with a $\sim0.5\%$ error in $d$, or $\sim1\%$ error in $\omega$. In the case of the radial modes of a linear Paul trap, this corresponds to a 2\% error in the ratio between the longitudinal and radial trapping frequencies. 

\textit{Imperfect pulses:} Throughout the analysis of gate fidelities within this work we assumed that each $\pi$-pulse is perfect. In any experimental demonstration the $\pi$-pulse fidelity would be an important contributing factor to the overall fidelity. The worst case impact of imperfect $\pi$-pulses can be shown to result in a reduction in fidelity that scales linearly with the number of $\pi$-pulses. This would give the overall fidelity $\text{F}_{\text{real}}$ as

\begin{eqnarray}
	 \text{F}_{\text{real}} \approx \left( 1 - 2 N_p \varepsilon \right) \text{F}_0,
\end{eqnarray}

where $\varepsilon$ is the population transfer error for each $\pi$-pulse, $N_p$ is the total number of $\pi$-pulses pairs in the gate sequence, and $\text{F}_{0}$ is the gate fidelity assuming perfect $\pi$-pulses. A detailed analysis of this error can be found here \cite{Gale2019}. The gates presented in this paper required between 50 (Fig.~\ref{fig:realistic_reps}) and 1000 $\pi$-pulses. Therefore, to achieve the fault tolerant threshold infidelity of $2 \times 10^{-4}$, a $\pi$-pulse error between $2 \times 10^{-6}$ and $10^{-8}$ would be necessary.  Typical intensity stability of pulsed lasers is lower than this, but this only corresponds to pulse error when using square pulses.  There are well-explored shaped pulses or multi-pulse schemes that correct the population transfer errors to first order \cite{Kabytayev2014,Vandersypen2005}.  Implementing them will be an extra experimental hurdle, but thereafter pulse errors should not limit the fast gates.

\begin{figure}
\includegraphics[width=0.9 \columnwidth]{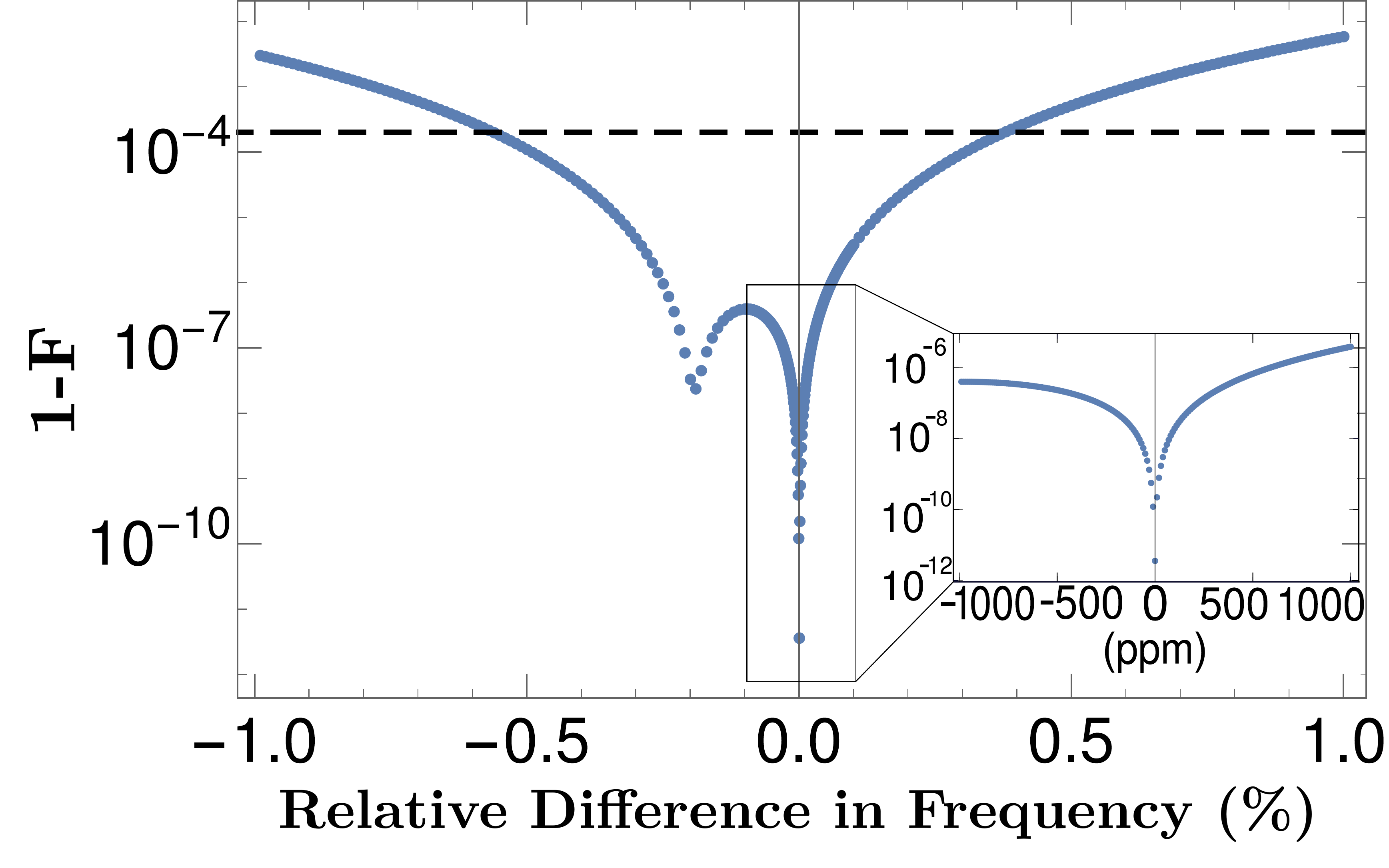}
\caption{The gate infidelity shown as a function of the difference in the secular trapping frequency between two harmonic trapping potentials, shown as a percentage difference in the main plot and a parts per million (ppm) difference in the inset.  }
\label{fig:sec_error} 
\end{figure}

\textit{Stray fields:} Finally, we consider the effects of stray fields on gate fidelities. Assuming the stray fields vary slowly with respect to the RF drive frequency, this error will result in a relative difference in the secular trapping frequencies between the two microtraps.  This will be linearly proportional to the relative intensity of the stray field and the trapping fields. We simulate this error by applying a frequency offset to one of the trapping potentials, and solving for the classical trajectories in phase space when the gate is applied. This is shown in Fig.~\ref{fig:sec_error}, which shows that although the difference between the breathing mode and the common motional mode in a microtrap architecture is small ($\sim10^{-4}$), correspondingly larger errors in the trapping frequencies ($\sim10^{-3}$) of the individual traps do not result in significant fidelity loss. The relationship between a static stray field and the secular trapping frequency will be $\frac{\Delta \omega}{\omega} \sim \sqrt{2} \frac{\Delta E }{ E}$. To conduct high fidelity gates it is necessary to ensure that stray fields are no larger than $0.1$-$0.2\%$ of the applied voltages used to generate the trapping potential.

\section{Conclusion}

We conclude that the micromotion present in ion trap experiments, either in microtraps or in linear Paul traps, may be harnessed to enhance the gate times and infidelities of fast gates using ultrafast pulses. We have presented a technique to implement this enhancement by locking the RF drive and repetition rate, and have also presented the conditions under which repetition rates faster than the RF drive may be used with high fidelity. We have further shown that this method remains robust to realistic experimental errors that would be encountered when implementing such fast gates.  Even in parameter regimes where the inclusion of micromotion makes only a small improvement, performing a high-fidelity fast gate requires it to be taken into account.  Implementing the gates presented in this work requires considerable technical control over the ultrafast pulses, which will be made significantly easier as schemes are found with fewer required $\pi$-pulse pairs.

\begin{acknowledgments}
This research was undertaken with the assistance of resources and services from the National Computational Infrastructure (NCI), which is supported by the Australian Government.
\end{acknowledgments}

\appendix

\section{\label{sec:opt}Fidelity Calculations in the absence of micromotion}

The material presented in this section here has been previously published in the supplementary material provided with \cite{Ratcliffe2018} and is reproduced here with permission for ease of reading.

We use numerical searches to find pulse timings that produce high quality gate operations, with the state-averaged fidelity $\text{F}$, given as the integral of the square of the norm of the overlap between the post-gate state with the target state integrated over all initial states. This is efficient to compute and it is strongly related to other distance measures for high-fidelity gates. As we examine fidelities extremely close to unity, we report the infidelity $1-\text{F}$.  This is a function of the phase mismatch $\Delta \phi$ around the target $\pi/4$ phase, and the phase space displacement of the motional modes $\Delta {P_p}$ given without micromotion as
\begin{eqnarray} 
    \Delta P_p = 2 \sqrt{\frac{\omega}{\omega_p}} \sum_{k}{z_k \sin{  (\omega_p t_k)}} \nonumber \\
    \Delta\phi = \left\vert \sum_p{8 \eta^2 \frac{\omega}{\omega_p} b_p^1 b_p^2 \sum_{i\neq j}{z_i z_j \sin{(\omega_p \lvert t_i - t_j \rvert )}}} \right\vert - \frac{\pi}{4}. \nonumber \\
\end{eqnarray}

For efficient computation of two-ion gates, we further simplify this measure by using a truncated expansion of the infidelity in these variables:. Assuming a thermal product state the infidelity is then given by
\begin{eqnarray}
	1 - \text{F} &\approx& \frac{2}{3} \Delta\phi^2 + \frac{4}{3} \sum_p (\frac{1}{2} + \overline{n}_p) ((b_p^1)^2 + (b_p^2)^2) \Delta {P_p}^2, \nonumber \\
    \label{costFunb}
\end{eqnarray}

where $\overline{n}_p$ is the mean motional occupation of the $p^\text{th}$ mode. While this approximate form is efficient for generating gate schemes, we use the full form when reporting achievable fidelities, for example, in the presence of multiple ions. We can see from Eq.~\eqref{costFunb} that the infidelity for a two-ion system, $1- F$, depends on the Lamb-Dicke parameter $\eta$, the angular frequencies collective motional modes $\omega_n$, the coupling of the $k$-th ion to the $p$-th mode, $b_p^k$, and the number of pulses in the $i$-th pulse train $z_i$.  The collective mode frequencies $\omega_n$ can be calculated from the mass of the ions $M$, the separation of the microtraps $d$, and the trapping frequency $\omega$ of the individual microtraps. \\
\\
We search for pulse timings that produce optimal gate fidelity within a given time bound. This optimisation is run as a set of local gradient searches in the three-dimensional parameter space of the pulse timings, over a large set of initial gate sequences. The highest fidelity of these local optimisations is then taken to be the optimal gate for that cap in the gate time. Note that the optimal gate occasionally takes less time than the maximum allowed.  By increasing the cap in total gate time and repeating this process, we map out the optimal fidelity for fast gates as a function of gate time.

\section{\label{sec:param}Key Parameter for Characterising Traps}

The material presented in this section here has been previously published in the supplementary material provided with \cite{Ratcliffe2018} and is reproduced here with permission for ease of reading.

We see from Eq.~\eqref{costFunb} that the system behaviour depends on the ratios of the frequencies of the collective modes. These are in turn functions of the geometry, and the dimensionless parameter $\xi = \frac{d^3\omega^2}{\alpha}$, where $\alpha = \frac{e^2}{4 \pi \varepsilon_0} \frac{1}{M}$. Here $e$ is the electron charge, $M$ the mass of the ions, and $\varepsilon_0$ the vacuum permittivity.  

For a two-ion system, there is only one ratio, so it entirely characterises the behaviour.    We define $\chi$ as the normalised difference between the breathing mode frequency and the common motional mode frequency $\chi=\frac{\omega_{\text{BR}} - \omega}{\omega}$ which can be expressed in terms of the more fundamental parameter $\xi$ as following:

\begin{eqnarray}
	\chi &=& \sqrt{\frac{1}{3} (9-\beta\gamma^{\frac{1}{3}}+\beta\gamma^{\frac{2}{3}}) }-1 
    \label{eq:chi}
\end{eqnarray}
where
\begin{eqnarray}
	\gamma = 1 + \frac{3(9+\sqrt{3}\sqrt{27+2\xi})}{\xi} \nonumber
\end{eqnarray}
and 
\begin{eqnarray}
	\beta = 9-\sqrt{3}\sqrt{27+2\xi} \nonumber
\end{eqnarray}
Its value lies in the range between 0 and $\sqrt{3}-1$.  The upper bound corresponds to the limit where both microtraps are merged, which is the case for standard linear trap geometries. \\
\\
When gates are conducted using the radial modes of ions co-trapped in a linear Paul trap, the relevant ratio is that between the rocking mode frequency $\omega_{R}$ and the common motional mode frequency in the transverse axis $\omega_T$. Taking the ratio of the axial and radial trapping frequencies to be $\kappa=\omega_A/\omega_T$, the rocking mode frequency will then be $\omega_R=(1-\kappa)\omega_T$. The parameter $\chi$ will then be given by
\begin{eqnarray}
	\chi &=& \sqrt{1-\kappa^2}-1.
    \label{eq:chiPaul}
\end{eqnarray}

In this case the value of $\chi$ will be between $-1$ and 0. The negative value of $\chi$ indicates a negative rate of phase acquisition compared to the microtrap system or the longitudinal modes of a linear Paul trap.\\
\\
Even for three or more ions, the system is still well characterised by $\chi$, which is the normalised gap to the lowest energy excitation in the system.  This is because it defines the rate of relative acquisition of phase between the excited and unexcited modes. 

\begin{widetext}
\section{\label{sec:mic}Fidelity Calculations in the presence of micromotion}

To determine the correct expression for fidelity we start by deriving the positions and velocities of the ions under a set of initial conditions. We do this for a general set of modes, defined by frequencies $\omega$, which makes this derivation suitable for both microtraps and radial kicks for ions in a common linear Paul trap. We then use this to derive an expression for the phase acquired at the end of a series of instantaneous momentum kicks. The displacement from equilibrium of a single ion with initial position and velocity $x_0$ and $v_0$ respectively at time $\tau=0$ in dimensionless time will be given by

\begin{eqnarray}
	x_{i,p}(\tau) &=& \frac{2 \beta}{\rho}  v_0 \frac{\omega}{\omega_p} \left(\sin (2 \pi \frac{\omega_p}{\omega} \tau) \left( \sigma_c f_c(\tau) + \sigma_s f_s(\tau) \right)+ \cos (2 \pi \frac{\omega_p}{\omega} \tau ) \left( \sigma_c f_s(\tau)-\sigma_s f_c(\tau) \right)\right) + \nonumber \\
	&~& \frac{2 \beta}{\rho} 4 \pi  x_0 \frac{\omega}{\omega_p} \sin (2 \pi \frac{\omega_p}{\omega} \tau) \left(\beta \sigma_s f_c(\tau)+2 f_c(\tau) \zeta_s-\beta  \sigma_c f_s(\tau)-2  f_s(\tau) \zeta_c \right) + \nonumber \\ 
	&~&  \frac{2 \beta}{\rho} 4 \pi  x_0 \frac{\omega}{\omega_p} \cos (2 \pi \frac{\omega_p}{\omega} \tau ) \left(\beta  \sigma_c f_c(\tau) +2 f_c(\tau) \zeta_c + \beta \sigma_s f_s(\tau) + 2 f_s(\tau) \zeta_s \right)
	\label{fsb}
\end{eqnarray}

where $x_0$ and $v_0$ have the same units as dimensionless time is being used, 

\begin{eqnarray}
f_c(\tau) &=&  \sum _{j=-\infty }^{\infty } C_j \cos \left(\frac{4 j \pi  \tau }{\beta }+j \phi_{\text{RF}} \right) \nonumber \\
f_s(\tau) &=&  \sum _{j=-\infty }^{\infty } C_j \sin \left(\frac{4 j \pi  \tau }{\beta }+j \phi_{\text{RF}} \right) \nonumber \\
\sigma_c &=& \sum _{j=-\infty }^{\infty } C_j \cos \left(j \phi_{\text{RF}} \right) \nonumber \\
\sigma_s &=& \sum _{j=-\infty }^{\infty } C_j \sin \left(j \phi_{\text{RF}} \right) \nonumber \\
\zeta_c &=& \sum _{j=-\infty }^{\infty } j C_j \cos \left(j \phi_{\text{RF}} \right) \nonumber \\
\zeta_s &=& \sum _{j=-\infty }^{\infty } j C_j \sin \left(j \phi_{\text{RF}} \right) \nonumber \\
\rho &=& {4 \pi \left( \sigma_c \left(\beta  \sigma_c + 2 \zeta_c \right) +\sigma_s \left(\beta  \sigma_s + 2 \zeta_s \right)\right)},
\end{eqnarray}
and the coefficients $C_j$ are given through the recurrence relationship shown below, obtained by substituting the solution into the ODE governing the motion.

\begin{eqnarray}
&& C_{j+1}-D_j C_j + C_{j-1} = 0 \nonumber \\
&& D_j = (a_x - (2j + \beta)^2)/q_x
\end{eqnarray}

By applying all pulses at the same fixed phase $\phi_{\text{RF}}$ in the micromotion cycle, the micromotion terms can be taken as a constant factor in the expression with $f_c(\tau) = \sigma_c$ and $f_s(\tau) = \sigma_s$. When we then consider the effect of an instantaneous pulse group imparting momentum $v_0$ at time $\tau=0$ the expression for the following motion will be given by

%Because the last line eqn becomes frac{2 \beta}{\rho} 4 \pi  x_0 \cos (2 \pi \frac{\omega_p}{\omega} \tau ) \left(\beta  \sigma_c^2 + 2 \sigma_c \sum _{n=-\infty }^{\infty } n D_n \cos \left(n \phi \right) + \beta \sigma_s^2 + 2 \sigma_s \sum _{n=-\infty }^{\infty } n D_n \sin \left(n \phi\right) \right) but the \rho cancels part of it

\begin{eqnarray}
	x_{i,p}(\tau) &=& \frac{2 \beta}{\rho} \left( v_0 \left( \sigma_c^2 + \sigma_s^2 \right) +  4 \pi  x_0 \left(2 \sigma_c \zeta_s - 2  \sigma_s \zeta_c \right) \right) \sin (2 \pi \frac{\omega_p}{\omega} \tau) + 8 \beta \pi  x_0 \cos (2 \pi \frac{\omega_p}{\omega} \tau )
\end{eqnarray}

and the velocity

\begin{eqnarray}
v_{i,p}(\tau) &=& \frac{4 \pi }{\rho } v_0 \cos (2 \pi  \tau ) \left( \sigma_c \left(\beta  \sigma_c + 2 \zeta_c \right) +\sigma_s \left(\beta  \sigma_s + 2 \zeta_s \right)\right)+ \nonumber \\
&~& \frac{4 \pi }{\rho } v_0 \sin (2 \pi  \tau ) \left(\sigma _s \zeta_c-\sigma _c \zeta_s\right) - \nonumber \\
&~& \frac{4 \pi }{\rho} 2 \pi  \beta  x_0 \frac{\omega}{\omega_p} \sin (2 \pi  \tau ) \left(4 \beta  \sigma _c \zeta_c +4 \beta  \sigma _s \zeta_s+4 \left(\zeta_s^2+\zeta_c^2\right)+\beta ^2 \sigma _c^2+\beta ^2 \sigma _s^2\right)
\end{eqnarray}

After a series of instantaneous momentum kicks $\lbrace P_j \rbrace$ at times $\lbrace \tau_j \rbrace$ the position after the $m^\text{th}$ kick will be given by

\begin{eqnarray}
	x_{i,p,m}(\tau) &=& \frac{2 \beta \left( \sigma_c^2 + \sigma_s^2  \right)}{\rho} \frac{\omega}{\omega_p} \sum_{j=1}^m P_j \sin (2 \pi \frac{\omega_p}{\omega} (\tau-\tau_j)) \nonumber \\
\end{eqnarray}

and the velocity will be given by

\begin{eqnarray}
	v_{i,p,m}(\tau) &=& \frac{4 \pi }{\rho }  \left(2 \sigma _c \zeta_c + 2 \sigma _s \zeta_s + \beta  \sigma _c^2+\beta  \sigma _s^2\right) \sum_{j=1}^m P_j \cos (2 \pi \frac{\omega_p}{\omega} (\tau-\tau_j) ) +  \frac{4 \pi }{\rho } \left(\sigma _s \zeta_c-\sigma _c \zeta_s\right)  \sum_{j=1}^m P_j\sin (2 \pi \frac{\omega_p}{\omega} (\tau-\tau_j) ) \nonumber \\
	&=& \sum_{j=1}^m P_j \cos (2 \pi \frac{\omega_p}{\omega} (\tau-\tau_j) ) +  \frac{4 \pi }{\rho } \left(\sigma _s \zeta_c-\sigma _c \zeta_s\right)  \sum_{j=1}^m P_j\sin (2 \pi \frac{\omega_p}{\omega} (\tau-\tau_j) ) 
\end{eqnarray}

The geometric phase from a sequence of $n$ pulses will be given by

\begin{eqnarray}
	\xi &=& \sum_k^{n-1} \sum_{i,p} (x_{i,p,k}(\tau_{k+1}) v_{i,p,k}(\tau_{k+1}) - x_{i,p,k}(\tau_{k}) v_{i,p,k}(\tau_{k})) \nonumber \\
	&=& \frac{2 \beta \left( \sigma_c^2 + \sigma_s^2  \right)}{\rho} \sum_k^{n-1} \sum_{i,p} \left( \frac{\omega}{\omega_p} \sum_{j=1}^k P_j \sin (2 \pi \frac{\omega_p}{\omega} (\tau_{k+1}-\tau_j)) \sum_{j=1}^k P_j \cos (2 \pi \frac{\omega_p}{\omega} (\tau_{k+1}-\tau_j) ) \right. \nonumber \\
	&-&  \frac{\omega}{\omega_p} \sum_{j=1}^k P_j \sin (2 \pi \frac{\omega_p}{\omega} (\tau_{k}-\tau_j)) \sum_{j=1}^k P_j \cos (2 \pi \frac{\omega_p}{\omega} (\tau_{k}-\tau_j) ) \nonumber \\
	&+& \frac{4 \pi }{\rho } \left(\sigma _s \zeta_c-\sigma _c \zeta_s\right)  \frac{\omega}{\omega_p} \left. \left( ( \sum_{j=1}^m P_j \sin (2 \pi \frac{\omega_p}{\omega} (\tau_{k+1}-\tau_j)))^2 - ( \sum_{j=1}^m P_j \sin (2 \pi \frac{\omega_p}{\omega} (\tau_{k}-\tau_j)))^2 \right) \right) 
\end{eqnarray}

The double sine terms can then be shown to be equivalent to the motional restoration term. This term will be small as part of our optimisation objective is to obtain motional restoration, and thus can be neglected. Giving the expression for $\xi$ as

\begin{eqnarray}
	\xi = \frac{2 \beta \left( \sigma_c^2 + \sigma_s^2  \right)}{\rho} \sum_k^{n-1} \sum_{i,p} &~&\left( \frac{\omega}{\omega_p} \sum_{j=1}^k P_j \sin (2 \pi \frac{\omega_p}{\omega} (\tau_{k+1}-\tau_j)) \sum_{j=1}^k P_j \cos (2 \pi \frac{\omega_p}{\omega} (\tau_{k+1}-\tau_j) ) \right. + \nonumber \\
	&~& \left. \frac{\omega}{\omega_p} \sum_{j=1}^k P_j \sin (2 \pi \frac{\omega_p}{\omega} (\tau_{k}-\tau_j)) \sum_{j=1}^k P_j \cos (2 \pi \frac{\omega_p}{\omega} (\tau_{k}-\tau_j) ) \right)
\end{eqnarray}

We denote the phase expression for an equivalent system without micromotion as $\xi_0$, given as 

\begin{eqnarray}
	\xi_0 = \sum_k^{n-1} \sum_{i,p} &~&\left( \frac{\omega}{\omega_p} \sum_{j=1}^k P_j \sin (2 \pi \frac{\omega_p}{\omega} (\tau_{k+1}-\tau_j)) \sum_{j=1}^k P_j \cos (2 \pi \frac{\omega_p}{\omega} (\tau_{k+1}-\tau_j) ) \right. + \nonumber \\
	&~& \left. \frac{\omega}{\omega_p} \sum_{j=1}^k P_j \sin (2 \pi \frac{\omega_p}{\omega} (\tau_{k}-\tau_j)) \sum_{j=1}^k P_j \cos (2 \pi \frac{\omega_p}{\omega} (\tau_{k}-\tau_j) ) \right)
\end{eqnarray}

the phase with micromotion can now be expressed as

\begin{eqnarray}
	\xi &=& \frac{\beta \left( \sigma_c^2 + \sigma_s^2  \right)}{\left( \sigma_c \left(\beta  \sigma_c + 2 \zeta_c \right) +\sigma_s \left(\beta  \sigma_s + 2 \zeta_s \right)\right)} \xi_0.
\end{eqnarray}

We can then define the term $\mu$ giving the increase in the maximum displacement $D'$ of the ion within the micromotion system after an instantaneous pulse, scaled to the maximum displacement of the same system without micromotion $D$

\begin{eqnarray}
	\mu &=& \frac{D' - D}{D} = \frac{\beta \left( \sigma_c^2 + \sigma_s^2  \right)}{\left( \sigma_c \left(\beta  \sigma_c + 2 \zeta_c \right) +\sigma_s \left(\beta  \sigma_s + 2 \zeta_s \right)\right)} \nonumber \\
\end{eqnarray}

giving the phase expression as
\begin{eqnarray}
	\xi &=& \mu \xi_0
\end{eqnarray}

The infidelity expression $1-F$, the motional restoration of the $p^{\text{th}}$ mode $\Delta P_p$ and the phase error $\Delta \phi$ will be given as
\begin{eqnarray} 
    1 - F &\approx& \frac{2}{3} \Delta\phi^2 + \frac{4}{3} \sum_p (\frac{1}{2} + \overline{n}_p) ((b_p^1)^2 + (b_p^2)^2) \Delta {P_p}^2 \nonumber \\
    \Delta P_p &=& 2 \mu \sqrt{\frac{\omega}{\omega_p}} \sum_{k}{z_k \sin{  (\omega_p t_k)}} \nonumber \\
    \Delta\phi &=& \left\vert \sum_p{8 \eta^2 \mu \frac{\omega}{\omega_p} b_p^1 b_p^2 \sum_{i\neq j}{z_i z_j \sin{(\omega_p \lvert t_i - t_j \rvert )}}} \right\vert - \frac{\pi}{4} \nonumber \\
\end{eqnarray}
\end{widetext}

\bibliographystyle{bibsty}
\bibliography{Micromotion2}

\begin{thebibliography}{48}%
\makeatletter
\providecommand \@ifxundefined [1]{%
 \@ifx{#1\undefined}
}%
\providecommand \@ifnum [1]{%
 \ifnum #1\expandafter \@firstoftwo
 \else \expandafter \@secondoftwo
 \fi
}%
\providecommand \@ifx [1]{%
 \ifx #1\expandafter \@firstoftwo
 \else \expandafter \@secondoftwo
 \fi
}%
\providecommand \natexlab [1]{#1}%
\providecommand \enquote  [1]{``#1''}%
\providecommand \bibnamefont  [1]{#1}%
\providecommand \bibfnamefont [1]{#1}%
\providecommand \citenamefont [1]{#1}%
\providecommand \href@noop [0]{\@secondoftwo}%
\providecommand \href [0]{\begingroup \@sanitize@url \@href}%
\providecommand \@href[1]{\@@startlink{#1}\@@href}%
\providecommand \@@href[1]{\endgroup#1\@@endlink}%
\providecommand \@sanitize@url [0]{\catcode `\\12\catcode `\$12\catcode
  `\&12\catcode `\#12\catcode `\^12\catcode `\_12\catcode `\%12\relax}%
\providecommand \@@startlink[1]{}%
\providecommand \@@endlink[0]{}%
\providecommand \url  [0]{\begingroup\@sanitize@url \@url }%
\providecommand \@url [1]{\endgroup\@href {#1}{\urlprefix }}%
\providecommand \urlprefix  [0]{URL }%
\providecommand \Eprint [0]{\href }%
\providecommand \doibase [0]{http://dx.doi.org/}%
\providecommand \selectlanguage [0]{\@gobble}%
\providecommand \bibinfo  [0]{\@secondoftwo}%
\providecommand \bibfield  [0]{\@secondoftwo}%
\providecommand \translation [1]{[#1]}%
\providecommand \BibitemOpen [0]{}%
\providecommand \bibitemStop [0]{}%
\providecommand \bibitemNoStop [0]{.\EOS\space}%
\providecommand \EOS [0]{\spacefactor3000\relax}%
\providecommand \BibitemShut  [1]{\csname bibitem#1\endcsname}%
\let\auto@bib@innerbib\@empty
%</preamble>
\bibitem [{\citenamefont {Nielsen}\ and\ \citenamefont
  {Chuang}(2010)}]{Nielsen2010}%
  \BibitemOpen
  \bibfield  {author} {\bibinfo {author} {\bibfnamefont {M.~A.}\ \bibnamefont
  {Nielsen}}\ and\ \bibinfo {author} {\bibfnamefont {I.~L.}\ \bibnamefont
  {Chuang}},\ }\href {\doibase 10.1017/CBO9780511976667} {\emph {\bibinfo
  {title} {Cambridge University Press}}}\ (\bibinfo  {publisher} {Cambridge
  University Press},\ \bibinfo {year} {2010})\ p.\ \bibinfo {pages}
  {702}\BibitemShut {NoStop}%
\bibitem [{\citenamefont {Makhlin}\ \emph {et~al.}(1999)\citenamefont
  {Makhlin}, \citenamefont {Sc{\"{o}}hn},\ and\ \citenamefont
  {Shnirman}}]{Makhlin1999}%
  \BibitemOpen
  \bibfield  {author} {\bibinfo {author} {\bibfnamefont {Y.}~\bibnamefont
  {Makhlin}}, \bibinfo {author} {\bibfnamefont {G.}~\bibnamefont
  {Sc{\"{o}}hn}}, \ and\ \bibinfo {author} {\bibfnamefont {A.}~\bibnamefont
  {Shnirman}},\ }\bibfield  {title} {\enquote {\bibinfo {title}
  {{Josephson-junction qubits with controlled couplings}},}\ }\href {\doibase
  10.1038/18613} {\bibfield  {journal} {\bibinfo  {journal} {Nature}\ }\textbf
  {\bibinfo {volume} {398}},\ \bibinfo {pages} {305} (\bibinfo {year}
  {1999})}\BibitemShut {NoStop}%
\bibitem [{\citenamefont {Neumann}\ \emph {et~al.}(2008)\citenamefont
  {Neumann}, \citenamefont {Mizuochi}, \citenamefont {Rempp}, \citenamefont
  {Hemmer}, \citenamefont {Watanabe}, \citenamefont {Yamasaki}, \citenamefont
  {Jacques}, \citenamefont {Gaebel}, \citenamefont {Jelezko},\ and\
  \citenamefont {Wrachtrup}}]{Neumann2008}%
  \BibitemOpen
  \bibfield  {author} {\bibinfo {author} {\bibfnamefont {P.}~\bibnamefont
  {Neumann}}, \bibinfo {author} {\bibfnamefont {N.}~\bibnamefont {Mizuochi}},
  \bibinfo {author} {\bibfnamefont {F.}~\bibnamefont {Rempp}}, \bibinfo
  {author} {\bibfnamefont {P.}~\bibnamefont {Hemmer}}, \bibinfo {author}
  {\bibfnamefont {H.}~\bibnamefont {Watanabe}}, \bibinfo {author}
  {\bibfnamefont {S.}~\bibnamefont {Yamasaki}}, \bibinfo {author}
  {\bibfnamefont {V.}~\bibnamefont {Jacques}}, \bibinfo {author} {\bibfnamefont
  {T.}~\bibnamefont {Gaebel}}, \bibinfo {author} {\bibfnamefont
  {F.}~\bibnamefont {Jelezko}}, \ and\ \bibinfo {author} {\bibfnamefont
  {J.}~\bibnamefont {Wrachtrup}},\ }\bibfield  {title} {\enquote {\bibinfo
  {title} {{Multipartite Entanglement Among Single Spins in Diamond}},}\ }\href
  {\doibase 10.1126/science.1157233} {\bibfield  {journal} {\bibinfo  {journal}
  {Science}\ }\textbf {\bibinfo {volume} {320}},\ \bibinfo {pages} {1326}
  (\bibinfo {year} {2008})}\BibitemShut {NoStop}%
\bibitem [{\citenamefont {Knill}\ \emph {et~al.}(2001)\citenamefont {Knill},
  \citenamefont {Laflamme},\ and\ \citenamefont {Milburn}}]{Knill2001}%
  \BibitemOpen
  \bibfield  {author} {\bibinfo {author} {\bibfnamefont {E.}~\bibnamefont
  {Knill}}, \bibinfo {author} {\bibfnamefont {R.}~\bibnamefont {Laflamme}}, \
  and\ \bibinfo {author} {\bibfnamefont {G.~J.}\ \bibnamefont {Milburn}},\
  }\bibfield  {title} {\enquote {\bibinfo {title} {{A scheme for efficient
  quantum computation with linear optics}},}\ }\href {\doibase
  10.1038/35051009} {\bibfield  {journal} {\bibinfo  {journal} {Nature}\
  }\textbf {\bibinfo {volume} {409}},\ \bibinfo {pages} {46} (\bibinfo {year}
  {2001})}\BibitemShut {NoStop}%
\bibitem [{\citenamefont {Cory}\ \emph {et~al.}(1997)\citenamefont {Cory},
  \citenamefont {Fahmy},\ and\ \citenamefont {Havel}}]{Cory1997}%
  \BibitemOpen
  \bibfield  {author} {\bibinfo {author} {\bibfnamefont {D.~G.}\ \bibnamefont
  {Cory}}, \bibinfo {author} {\bibfnamefont {A.~F.}\ \bibnamefont {Fahmy}}, \
  and\ \bibinfo {author} {\bibfnamefont {T.~F.}\ \bibnamefont {Havel}},\
  }\bibfield  {title} {\enquote {\bibinfo {title} {{Ensemble quantum computing
  by NMR spectroscopy}},}\ }\href {\doibase 10.1073/pnas.94.5.1634} {\bibfield
  {journal} {\bibinfo  {journal} {Proceedings of the National Academy of
  Sciences}\ }\textbf {\bibinfo {volume} {94}},\ \bibinfo {pages} {1634}
  (\bibinfo {year} {1997})}\BibitemShut {NoStop}%
\bibitem [{\citenamefont {Nayak}\ \emph {et~al.}(2008)\citenamefont {Nayak},
  \citenamefont {Simon}, \citenamefont {Stern}, \citenamefont {Freedman},\ and\
  \citenamefont {{Das Sarma}}}]{Nayak2008}%
  \BibitemOpen
  \bibfield  {author} {\bibinfo {author} {\bibfnamefont {C.}~\bibnamefont
  {Nayak}}, \bibinfo {author} {\bibfnamefont {S.~H.}\ \bibnamefont {Simon}},
  \bibinfo {author} {\bibfnamefont {A.}~\bibnamefont {Stern}}, \bibinfo
  {author} {\bibfnamefont {M.}~\bibnamefont {Freedman}}, \ and\ \bibinfo
  {author} {\bibfnamefont {S.}~\bibnamefont {{Das Sarma}}},\ }\bibfield
  {title} {\enquote {\bibinfo {title} {{Non-Abelian anyons and topological
  quantum computation}},}\ }\href {\doibase 10.1103/RevModPhys.80.1083}
  {\bibfield  {journal} {\bibinfo  {journal} {Reviews of Modern Physics}\
  }\textbf {\bibinfo {volume} {80}},\ \bibinfo {pages} {1083} (\bibinfo {year}
  {2008})}\BibitemShut {NoStop}%
\bibitem [{\citenamefont {Loss}\ and\ \citenamefont
  {DiVincenzo}(1998)}]{Loss1998}%
  \BibitemOpen
  \bibfield  {author} {\bibinfo {author} {\bibfnamefont {D.}~\bibnamefont
  {Loss}}\ and\ \bibinfo {author} {\bibfnamefont {D.~P.}\ \bibnamefont
  {DiVincenzo}},\ }\bibfield  {title} {\enquote {\bibinfo {title} {{Quantum
  computation with quantum dots}},}\ }\href {\doibase 10.1103/PhysRevA.57.120}
  {\bibfield  {journal} {\bibinfo  {journal} {Physical Review A}\ }\textbf
  {\bibinfo {volume} {57}},\ \bibinfo {pages} {120} (\bibinfo {year}
  {1998})}\BibitemShut {NoStop}%
\bibitem [{\citenamefont {Kane}(1998)}]{Kane1998}%
  \BibitemOpen
  \bibfield  {author} {\bibinfo {author} {\bibfnamefont {B.~E.}\ \bibnamefont
  {Kane}},\ }\bibfield  {title} {\enquote {\bibinfo {title} {{A silicon-based
  nuclear spin quantum computer}},}\ }\href {\doibase 10.1038/30156} {\bibfield
   {journal} {\bibinfo  {journal} {Nature}\ }\textbf {\bibinfo {volume}
  {393}},\ \bibinfo {pages} {133} (\bibinfo {year} {1998})}\BibitemShut
  {NoStop}%
\bibitem [{\citenamefont {Fowler}\ \emph {et~al.}(2012)\citenamefont {Fowler},
  \citenamefont {Mariantoni}, \citenamefont {Martinis},\ and\ \citenamefont
  {Cleland}}]{Fowler2012}%
  \BibitemOpen
  \bibfield  {author} {\bibinfo {author} {\bibfnamefont {A.~G.}\ \bibnamefont
  {Fowler}}, \bibinfo {author} {\bibfnamefont {M.}~\bibnamefont {Mariantoni}},
  \bibinfo {author} {\bibfnamefont {J.~M.}\ \bibnamefont {Martinis}}, \ and\
  \bibinfo {author} {\bibfnamefont {A.~N.}\ \bibnamefont {Cleland}},\
  }\bibfield  {title} {\enquote {\bibinfo {title} {{Surface codes: Towards
  practical large-scale quantum computation}},}\ }\href {\doibase
  10.1103/PhysRevA.86.032324} {\bibfield  {journal} {\bibinfo  {journal}
  {Physical Review A}\ }\textbf {\bibinfo {volume} {86}},\ \bibinfo {pages}
  {032324} (\bibinfo {year} {2012})}\BibitemShut {NoStop}%
\bibitem [{\citenamefont {DiVincenzo}(2000)}]{DiVincenzo2000a}%
  \BibitemOpen
  \bibfield  {author} {\bibinfo {author} {\bibfnamefont {D.~P.}\ \bibnamefont
  {DiVincenzo}},\ }\bibfield  {title} {\enquote {\bibinfo {title} {{The
  Physical Implementation of Quantum Computation}},}\ }\href {\doibase
  10.1002/1521-3978(200009)48:9/11<771::AID-PROP771>3.0.CO;2-E} {\bibfield
  {journal} {\bibinfo  {journal} {Fortschritte der Physik}\ }\textbf {\bibinfo
  {volume} {48}},\ \bibinfo {pages} {771} (\bibinfo {year} {2000})}\BibitemShut
  {NoStop}%
\bibitem [{\citenamefont {Waldherr}\ \emph {et~al.}(2014)\citenamefont
  {Waldherr}, \citenamefont {Wang}, \citenamefont {Zaiser}, \citenamefont
  {Jamali}, \citenamefont {Schulte-Herbr{\"{u}}ggen}, \citenamefont {Abe},
  \citenamefont {Ohshima}, \citenamefont {Isoya}, \citenamefont {Du},
  \citenamefont {Neumann},\ and\ \citenamefont {Wrachtrup}}]{Waldherr2014}%
  \BibitemOpen
  \bibfield  {author} {\bibinfo {author} {\bibfnamefont {G.}~\bibnamefont
  {Waldherr}}, \bibinfo {author} {\bibfnamefont {Y.}~\bibnamefont {Wang}},
  \bibinfo {author} {\bibfnamefont {S.}~\bibnamefont {Zaiser}}, \bibinfo
  {author} {\bibfnamefont {M.}~\bibnamefont {Jamali}}, \bibinfo {author}
  {\bibfnamefont {T.}~\bibnamefont {Schulte-Herbr{\"{u}}ggen}}, \bibinfo
  {author} {\bibfnamefont {H.}~\bibnamefont {Abe}}, \bibinfo {author}
  {\bibfnamefont {T.}~\bibnamefont {Ohshima}}, \bibinfo {author} {\bibfnamefont
  {J.}~\bibnamefont {Isoya}}, \bibinfo {author} {\bibfnamefont {J.~F.}\
  \bibnamefont {Du}}, \bibinfo {author} {\bibfnamefont {P.}~\bibnamefont
  {Neumann}}, \ and\ \bibinfo {author} {\bibfnamefont {J.}~\bibnamefont
  {Wrachtrup}},\ }\bibfield  {title} {\enquote {\bibinfo {title} {{Quantum
  error correction in a solid-state hybrid spin register}},}\ }\href {\doibase
  10.1038/nature12919} {\bibfield  {journal} {\bibinfo  {journal} {Nature}\
  }\textbf {\bibinfo {volume} {506}},\ \bibinfo {pages} {204} (\bibinfo {year}
  {2014})}\BibitemShut {NoStop}%
\bibitem [{\citenamefont {Veldhorst}\ \emph {et~al.}(2014)\citenamefont
  {Veldhorst}, \citenamefont {Hwang}, \citenamefont {Yang}, \citenamefont
  {Leenstra}, \citenamefont {de~Ronde}, \citenamefont {Dehollain},
  \citenamefont {Muhonen}, \citenamefont {Hudson}, \citenamefont {Itoh},
  \citenamefont {Morello},\ and\ \citenamefont {Dzurak}}]{Veldhorst2014}%
  \BibitemOpen
  \bibfield  {author} {\bibinfo {author} {\bibfnamefont {M.}~\bibnamefont
  {Veldhorst}}, \bibinfo {author} {\bibfnamefont {J.~C.~C.}\ \bibnamefont
  {Hwang}}, \bibinfo {author} {\bibfnamefont {C.~H.}\ \bibnamefont {Yang}},
  \bibinfo {author} {\bibfnamefont {A.~W.}\ \bibnamefont {Leenstra}}, \bibinfo
  {author} {\bibfnamefont {B.}~\bibnamefont {de~Ronde}}, \bibinfo {author}
  {\bibfnamefont {J.~P.}\ \bibnamefont {Dehollain}}, \bibinfo {author}
  {\bibfnamefont {J.~T.}\ \bibnamefont {Muhonen}}, \bibinfo {author}
  {\bibfnamefont {F.~E.}\ \bibnamefont {Hudson}}, \bibinfo {author}
  {\bibfnamefont {K.~M.}\ \bibnamefont {Itoh}}, \bibinfo {author}
  {\bibfnamefont {A.}~\bibnamefont {Morello}}, \ and\ \bibinfo {author}
  {\bibfnamefont {A.~S.}\ \bibnamefont {Dzurak}},\ }\bibfield  {title}
  {\enquote {\bibinfo {title} {{An addressable quantum dot qubit with
  fault-tolerant control-fidelity}},}\ }\href {\doibase 10.1038/nnano.2014.216}
  {\bibfield  {journal} {\bibinfo  {journal} {Nature Nanotechnology}\ }\textbf
  {\bibinfo {volume} {9}},\ \bibinfo {pages} {981} (\bibinfo {year}
  {2014})}\BibitemShut {NoStop}%
\bibitem [{\citenamefont {Dolde}\ \emph {et~al.}(2014)\citenamefont {Dolde},
  \citenamefont {Bergholm}, \citenamefont {Wang}, \citenamefont {Jakobi},
  \citenamefont {Naydenov}, \citenamefont {Pezzagna}, \citenamefont {Meijer},
  \citenamefont {Jelezko}, \citenamefont {Neumann}, \citenamefont
  {Schulte-Herbr{\"{u}}ggen}, \citenamefont {Biamonte},\ and\ \citenamefont
  {Wrachtrup}}]{Dolde2014}%
  \BibitemOpen
  \bibfield  {author} {\bibinfo {author} {\bibfnamefont {F.}~\bibnamefont
  {Dolde}}, \bibinfo {author} {\bibfnamefont {V.}~\bibnamefont {Bergholm}},
  \bibinfo {author} {\bibfnamefont {Y.}~\bibnamefont {Wang}}, \bibinfo {author}
  {\bibfnamefont {I.}~\bibnamefont {Jakobi}}, \bibinfo {author} {\bibfnamefont
  {B.}~\bibnamefont {Naydenov}}, \bibinfo {author} {\bibfnamefont
  {S.}~\bibnamefont {Pezzagna}}, \bibinfo {author} {\bibfnamefont
  {J.}~\bibnamefont {Meijer}}, \bibinfo {author} {\bibfnamefont
  {F.}~\bibnamefont {Jelezko}}, \bibinfo {author} {\bibfnamefont
  {P.}~\bibnamefont {Neumann}}, \bibinfo {author} {\bibfnamefont
  {T.}~\bibnamefont {Schulte-Herbr{\"{u}}ggen}}, \bibinfo {author}
  {\bibfnamefont {J.}~\bibnamefont {Biamonte}}, \ and\ \bibinfo {author}
  {\bibfnamefont {J.}~\bibnamefont {Wrachtrup}},\ }\bibfield  {title} {\enquote
  {\bibinfo {title} {{High-fidelity spin entanglement using optimal
  control}},}\ }\href {\doibase 10.1038/ncomms4371} {\bibfield  {journal}
  {\bibinfo  {journal} {Nature Communications}\ }\textbf {\bibinfo {volume}
  {5}},\ \bibinfo {pages} {3371} (\bibinfo {year} {2014})}\BibitemShut
  {NoStop}%
\bibitem [{\citenamefont {Barends}\ \emph {et~al.}(2014)\citenamefont
  {Barends}, \citenamefont {Kelly}, \citenamefont {Megrant}, \citenamefont
  {Veitia}, \citenamefont {Sank}, \citenamefont {Jeffrey}, \citenamefont
  {White}, \citenamefont {Mutus}, \citenamefont {Fowler}, \citenamefont
  {Campbell}, \citenamefont {Chen}, \citenamefont {Chen}, \citenamefont
  {Chiaro}, \citenamefont {Dunsworth}, \citenamefont {Neill}, \citenamefont
  {O'Malley}, \citenamefont {Roushan}, \citenamefont {Vainsencher},
  \citenamefont {Wenner}, \citenamefont {Korotkov}, \citenamefont {Cleland},\
  and\ \citenamefont {Martinis}}]{Barends2014}%
  \BibitemOpen
  \bibfield  {author} {\bibinfo {author} {\bibfnamefont {R.}~\bibnamefont
  {Barends}}, \bibinfo {author} {\bibfnamefont {J.}~\bibnamefont {Kelly}},
  \bibinfo {author} {\bibfnamefont {A.}~\bibnamefont {Megrant}}, \bibinfo
  {author} {\bibfnamefont {A.}~\bibnamefont {Veitia}}, \bibinfo {author}
  {\bibfnamefont {D.}~\bibnamefont {Sank}}, \bibinfo {author} {\bibfnamefont
  {E.}~\bibnamefont {Jeffrey}}, \bibinfo {author} {\bibfnamefont {T.~C.}\
  \bibnamefont {White}}, \bibinfo {author} {\bibfnamefont {J.}~\bibnamefont
  {Mutus}}, \bibinfo {author} {\bibfnamefont {A.~G.}\ \bibnamefont {Fowler}},
  \bibinfo {author} {\bibfnamefont {B.}~\bibnamefont {Campbell}}, \bibinfo
  {author} {\bibfnamefont {Y.}~\bibnamefont {Chen}}, \bibinfo {author}
  {\bibfnamefont {Z.}~\bibnamefont {Chen}}, \bibinfo {author} {\bibfnamefont
  {B.}~\bibnamefont {Chiaro}}, \bibinfo {author} {\bibfnamefont
  {A.}~\bibnamefont {Dunsworth}}, \bibinfo {author} {\bibfnamefont
  {C.}~\bibnamefont {Neill}}, \bibinfo {author} {\bibfnamefont
  {P.}~\bibnamefont {O'Malley}}, \bibinfo {author} {\bibfnamefont
  {P.}~\bibnamefont {Roushan}}, \bibinfo {author} {\bibfnamefont
  {A.}~\bibnamefont {Vainsencher}}, \bibinfo {author} {\bibfnamefont
  {J.}~\bibnamefont {Wenner}}, \bibinfo {author} {\bibfnamefont {A.~N.}\
  \bibnamefont {Korotkov}}, \bibinfo {author} {\bibfnamefont {A.~N.}\
  \bibnamefont {Cleland}}, \ and\ \bibinfo {author} {\bibfnamefont {J.~M.}\
  \bibnamefont {Martinis}},\ }\bibfield  {title} {\enquote {\bibinfo {title}
  {{Superconducting quantum circuits at the surface code threshold for fault
  tolerance}},}\ }\href {\doibase 10.1038/nature13171} {\bibfield  {journal}
  {\bibinfo  {journal} {Nature}\ }\textbf {\bibinfo {volume} {508}},\ \bibinfo
  {pages} {500} (\bibinfo {year} {2014})}\BibitemShut {NoStop}%
\bibitem [{\citenamefont {Friis}\ \emph {et~al.}(2018)\citenamefont {Friis},
  \citenamefont {Marty}, \citenamefont {Maier}, \citenamefont {Hempel},
  \citenamefont {Holz{\"{a}}pfel}, \citenamefont {Jurcevic}, \citenamefont
  {Plenio}, \citenamefont {Huber}, \citenamefont {Roos}, \citenamefont
  {Blatt},\ and\ \citenamefont {Lanyon}}]{Friis2018}%
  \BibitemOpen
  \bibfield  {author} {\bibinfo {author} {\bibfnamefont {N.}~\bibnamefont
  {Friis}}, \bibinfo {author} {\bibfnamefont {O.}~\bibnamefont {Marty}},
  \bibinfo {author} {\bibfnamefont {C.}~\bibnamefont {Maier}}, \bibinfo
  {author} {\bibfnamefont {C.}~\bibnamefont {Hempel}}, \bibinfo {author}
  {\bibfnamefont {M.}~\bibnamefont {Holz{\"{a}}pfel}}, \bibinfo {author}
  {\bibfnamefont {P.}~\bibnamefont {Jurcevic}}, \bibinfo {author}
  {\bibfnamefont {M.~B.}\ \bibnamefont {Plenio}}, \bibinfo {author}
  {\bibfnamefont {M.}~\bibnamefont {Huber}}, \bibinfo {author} {\bibfnamefont
  {C.}~\bibnamefont {Roos}}, \bibinfo {author} {\bibfnamefont {R.}~\bibnamefont
  {Blatt}}, \ and\ \bibinfo {author} {\bibfnamefont {B.}~\bibnamefont
  {Lanyon}},\ }\bibfield  {title} {\enquote {\bibinfo {title} {{Observation of
  Entangled States of a Fully Controlled 20-Qubit System}},}\ }\href {\doibase
  10.1103/PhysRevX.8.021012} {\bibfield  {journal} {\bibinfo  {journal}
  {Physical Review X}\ }\textbf {\bibinfo {volume} {8}},\ \bibinfo {pages}
  {021012} (\bibinfo {year} {2018})}\BibitemShut {NoStop}%
\bibitem [{\citenamefont {Hempel}\ \emph {et~al.}(2018)\citenamefont {Hempel},
  \citenamefont {Maier}, \citenamefont {Romero}, \citenamefont {McClean},
  \citenamefont {Monz}, \citenamefont {Shen}, \citenamefont {Jurcevic},
  \citenamefont {Lanyon}, \citenamefont {Love}, \citenamefont {Babbush},
  \citenamefont {Aspuru-Guzik}, \citenamefont {Blatt},\ and\ \citenamefont
  {Roos}}]{Hempel2018}%
  \BibitemOpen
  \bibfield  {author} {\bibinfo {author} {\bibfnamefont {C.}~\bibnamefont
  {Hempel}}, \bibinfo {author} {\bibfnamefont {C.}~\bibnamefont {Maier}},
  \bibinfo {author} {\bibfnamefont {J.}~\bibnamefont {Romero}}, \bibinfo
  {author} {\bibfnamefont {J.}~\bibnamefont {McClean}}, \bibinfo {author}
  {\bibfnamefont {T.}~\bibnamefont {Monz}}, \bibinfo {author} {\bibfnamefont
  {H.}~\bibnamefont {Shen}}, \bibinfo {author} {\bibfnamefont {P.}~\bibnamefont
  {Jurcevic}}, \bibinfo {author} {\bibfnamefont {B.~P.}\ \bibnamefont
  {Lanyon}}, \bibinfo {author} {\bibfnamefont {P.}~\bibnamefont {Love}},
  \bibinfo {author} {\bibfnamefont {R.}~\bibnamefont {Babbush}}, \bibinfo
  {author} {\bibfnamefont {A.}~\bibnamefont {Aspuru-Guzik}}, \bibinfo {author}
  {\bibfnamefont {R.}~\bibnamefont {Blatt}}, \ and\ \bibinfo {author}
  {\bibfnamefont {C.~F.}\ \bibnamefont {Roos}},\ }\bibfield  {title} {\enquote
  {\bibinfo {title} {{Quantum Chemistry Calculations on a Trapped-Ion Quantum
  Simulator}},}\ }\href {\doibase 10.1103/PhysRevX.8.031022} {\bibfield
  {journal} {\bibinfo  {journal} {Physical Review X}\ }\textbf {\bibinfo
  {volume} {8}},\ \bibinfo {pages} {031022} (\bibinfo {year}
  {2018})}\BibitemShut {NoStop}%
\bibitem [{\citenamefont {Nigg}\ \emph {et~al.}(2014)\citenamefont {Nigg},
  \citenamefont {Muller}, \citenamefont {Martinez}, \citenamefont {Schindler},
  \citenamefont {Hennrich}, \citenamefont {Monz}, \citenamefont
  {Martin-Delgado},\ and\ \citenamefont {Blatt}}]{Nigg2014}%
  \BibitemOpen
  \bibfield  {author} {\bibinfo {author} {\bibfnamefont {D.}~\bibnamefont
  {Nigg}}, \bibinfo {author} {\bibfnamefont {M.}~\bibnamefont {Muller}},
  \bibinfo {author} {\bibfnamefont {E.~A.}\ \bibnamefont {Martinez}}, \bibinfo
  {author} {\bibfnamefont {P.}~\bibnamefont {Schindler}}, \bibinfo {author}
  {\bibfnamefont {M.}~\bibnamefont {Hennrich}}, \bibinfo {author}
  {\bibfnamefont {T.}~\bibnamefont {Monz}}, \bibinfo {author} {\bibfnamefont
  {M.~A.}\ \bibnamefont {Martin-Delgado}}, \ and\ \bibinfo {author}
  {\bibfnamefont {R.}~\bibnamefont {Blatt}},\ }\bibfield  {title} {\enquote
  {\bibinfo {title} {{Quantum computations on a topologically encoded
  qubit}},}\ }\href {\doibase 10.1126/science.1253742} {\bibfield  {journal}
  {\bibinfo  {journal} {Science}\ }\textbf {\bibinfo {volume} {345}},\ \bibinfo
  {pages} {302} (\bibinfo {year} {2014})}\BibitemShut {NoStop}%
\bibitem [{\citenamefont {Zhang}\ \emph {et~al.}(2017)\citenamefont {Zhang},
  \citenamefont {Pagano}, \citenamefont {Hess}, \citenamefont {Kyprianidis},
  \citenamefont {Becker}, \citenamefont {Kaplan}, \citenamefont {Gorshkov},
  \citenamefont {Gong},\ and\ \citenamefont {Monroe}}]{Zhang2017}%
  \BibitemOpen
  \bibfield  {author} {\bibinfo {author} {\bibfnamefont {J.}~\bibnamefont
  {Zhang}}, \bibinfo {author} {\bibfnamefont {G.}~\bibnamefont {Pagano}},
  \bibinfo {author} {\bibfnamefont {P.~W.}\ \bibnamefont {Hess}}, \bibinfo
  {author} {\bibfnamefont {A.}~\bibnamefont {Kyprianidis}}, \bibinfo {author}
  {\bibfnamefont {P.}~\bibnamefont {Becker}}, \bibinfo {author} {\bibfnamefont
  {H.}~\bibnamefont {Kaplan}}, \bibinfo {author} {\bibfnamefont {A.~V.}\
  \bibnamefont {Gorshkov}}, \bibinfo {author} {\bibfnamefont {Z.-X.}\
  \bibnamefont {Gong}}, \ and\ \bibinfo {author} {\bibfnamefont
  {C.}~\bibnamefont {Monroe}},\ }\bibfield  {title} {\enquote {\bibinfo {title}
  {{Observation of a many-body dynamical phase transition with a 53-qubit
  quantum simulator}},}\ }\href {\doibase 10.1038/nature24654} {\bibfield
  {journal} {\bibinfo  {journal} {Nature}\ }\textbf {\bibinfo {volume} {551}},\
  \bibinfo {pages} {601} (\bibinfo {year} {2017})}\BibitemShut {NoStop}%
\bibitem [{\citenamefont {Brooks}\ and\ \citenamefont
  {Preskill}(2013)}]{Brooks2013}%
  \BibitemOpen
  \bibfield  {author} {\bibinfo {author} {\bibfnamefont {P.}~\bibnamefont
  {Brooks}}\ and\ \bibinfo {author} {\bibfnamefont {J.}~\bibnamefont
  {Preskill}},\ }\bibfield  {title} {\enquote {\bibinfo {title}
  {{Fault-tolerant quantum computation with asymmetric Bacon-Shor codes}},}\
  }\href {\doibase 10.1103/PhysRevA.87.032310} {\bibfield  {journal} {\bibinfo
  {journal} {Physical Review A}\ }\textbf {\bibinfo {volume} {87}},\ \bibinfo
  {pages} {032310} (\bibinfo {year} {2013})}\BibitemShut {NoStop}%
\bibitem [{\citenamefont {Ballance}\ \emph {et~al.}(2016)\citenamefont
  {Ballance}, \citenamefont {Harty}, \citenamefont {Linke}, \citenamefont
  {Sepiol},\ and\ \citenamefont {Lucas}}]{Ballance2016}%
  \BibitemOpen
  \bibfield  {author} {\bibinfo {author} {\bibfnamefont {C.~J.}\ \bibnamefont
  {Ballance}}, \bibinfo {author} {\bibfnamefont {T.~P.}\ \bibnamefont {Harty}},
  \bibinfo {author} {\bibfnamefont {N.~M.}\ \bibnamefont {Linke}}, \bibinfo
  {author} {\bibfnamefont {M.~A.}\ \bibnamefont {Sepiol}}, \ and\ \bibinfo
  {author} {\bibfnamefont {D.~M.}\ \bibnamefont {Lucas}},\ }\bibfield  {title}
  {\enquote {\bibinfo {title} {{High-Fidelity Quantum Logic Gates Using
  Trapped-Ion Hyperfine Qubits}},}\ }\href {\doibase
  10.1103/PhysRevLett.117.060504} {\bibfield  {journal} {\bibinfo  {journal}
  {Physical Review Letters}\ }\textbf {\bibinfo {volume} {117}},\ \bibinfo
  {pages} {060504} (\bibinfo {year} {2016})}\BibitemShut {NoStop}%
\bibitem [{\citenamefont {Gaebler}\ \emph {et~al.}(2016)\citenamefont
  {Gaebler}, \citenamefont {Tan}, \citenamefont {Lin}, \citenamefont {Wan},
  \citenamefont {Bowler}, \citenamefont {Keith}, \citenamefont {Glancy},
  \citenamefont {Coakley}, \citenamefont {Knill}, \citenamefont {Leibfried},\
  and\ \citenamefont {Wineland}}]{Gaebler2016}%
  \BibitemOpen
  \bibfield  {author} {\bibinfo {author} {\bibfnamefont {J.~P.}\ \bibnamefont
  {Gaebler}}, \bibinfo {author} {\bibfnamefont {T.~R.}\ \bibnamefont {Tan}},
  \bibinfo {author} {\bibfnamefont {Y.}~\bibnamefont {Lin}}, \bibinfo {author}
  {\bibfnamefont {Y.}~\bibnamefont {Wan}}, \bibinfo {author} {\bibfnamefont
  {R.}~\bibnamefont {Bowler}}, \bibinfo {author} {\bibfnamefont {A.~C.}\
  \bibnamefont {Keith}}, \bibinfo {author} {\bibfnamefont {S.}~\bibnamefont
  {Glancy}}, \bibinfo {author} {\bibfnamefont {K.}~\bibnamefont {Coakley}},
  \bibinfo {author} {\bibfnamefont {E.}~\bibnamefont {Knill}}, \bibinfo
  {author} {\bibfnamefont {D.}~\bibnamefont {Leibfried}}, \ and\ \bibinfo
  {author} {\bibfnamefont {D.~J.}\ \bibnamefont {Wineland}},\ }\bibfield
  {title} {\enquote {\bibinfo {title} {{High-Fidelity Universal Gate Set for
  {\$}{\^{}}{\{}9{\}}{\$}Be{\$}{\^{}}+{\$} Ion Qubits}},}\ }\href {\doibase
  10.1103/PhysRevLett.117.060505} {\bibfield  {journal} {\bibinfo  {journal}
  {Physical Review Letters}\ }\textbf {\bibinfo {volume} {117}},\ \bibinfo
  {pages} {060505} (\bibinfo {year} {2016})}\BibitemShut {NoStop}%
\bibitem [{\citenamefont {Baldwin}\ \emph {et~al.}(2019)\citenamefont
  {Baldwin}, \citenamefont {Bjork}, \citenamefont {Gaebler}, \citenamefont
  {Hayes},\ and\ \citenamefont {Stack}}]{Baldwin2019}%
  \BibitemOpen
  \bibfield  {author} {\bibinfo {author} {\bibfnamefont {C.~H.}\ \bibnamefont
  {Baldwin}}, \bibinfo {author} {\bibfnamefont {B.~J.}\ \bibnamefont {Bjork}},
  \bibinfo {author} {\bibfnamefont {J.~P.}\ \bibnamefont {Gaebler}}, \bibinfo
  {author} {\bibfnamefont {D.}~\bibnamefont {Hayes}}, \ and\ \bibinfo {author}
  {\bibfnamefont {D.}~\bibnamefont {Stack}},\ }\bibfield  {title} {\enquote
  {\bibinfo {title} {{Subspace benchmarking high-fidelity entangling operations
  with trapped ions}},}\ }\href {http://arxiv.org/abs/1911.00085} {\bibfield
  {journal} {\bibinfo  {journal} {arXiv preprint}\ } (\bibinfo {year}
  {2019})}\BibitemShut {NoStop}%
\bibitem [{\citenamefont {Casanova}\ \emph {et~al.}(2012)\citenamefont
  {Casanova}, \citenamefont {Mezzacapo}, \citenamefont {Lamata},\ and\
  \citenamefont {Solano}}]{Casanova2012}%
  \BibitemOpen
  \bibfield  {author} {\bibinfo {author} {\bibfnamefont {J.}~\bibnamefont
  {Casanova}}, \bibinfo {author} {\bibfnamefont {A.}~\bibnamefont {Mezzacapo}},
  \bibinfo {author} {\bibfnamefont {L.}~\bibnamefont {Lamata}}, \ and\ \bibinfo
  {author} {\bibfnamefont {E.}~\bibnamefont {Solano}},\ }\bibfield  {title}
  {\enquote {\bibinfo {title} {{Quantum Simulation of Interacting Fermion
  Lattice Models in Trapped Ions}},}\ }\href {\doibase
  10.1103/PhysRevLett.108.190502} {\bibfield  {journal} {\bibinfo  {journal}
  {Physical Review Letters}\ }\textbf {\bibinfo {volume} {108}},\ \bibinfo
  {pages} {190502} (\bibinfo {year} {2012})}\BibitemShut {NoStop}%
\bibitem [{\citenamefont {Garc{\'{i}}a-Ripoll}\ \emph
  {et~al.}(2005)\citenamefont {Garc{\'{i}}a-Ripoll}, \citenamefont {Zoller},\
  and\ \citenamefont {Cirac}}]{Garcia-Ripoll2004}%
  \BibitemOpen
  \bibfield  {author} {\bibinfo {author} {\bibfnamefont {J.~J.}\ \bibnamefont
  {Garc{\'{i}}a-Ripoll}}, \bibinfo {author} {\bibfnamefont {P.}~\bibnamefont
  {Zoller}}, \ and\ \bibinfo {author} {\bibfnamefont {J.~I.}\ \bibnamefont
  {Cirac}},\ }\bibfield  {title} {\enquote {\bibinfo {title} {{Coherent control
  of trapped ions using off-resonant lasers}},}\ }\href {\doibase
  10.1103/PhysRevA.71.062309} {\bibfield  {journal} {\bibinfo  {journal}
  {Physical Review A}\ }\textbf {\bibinfo {volume} {71}},\ \bibinfo {pages}
  {062309} (\bibinfo {year} {2005})}\BibitemShut {NoStop}%
\bibitem [{\citenamefont {Duan}(2004)}]{Duan2004}%
  \BibitemOpen
  \bibfield  {author} {\bibinfo {author} {\bibfnamefont {L.-M.}\ \bibnamefont
  {Duan}},\ }\bibfield  {title} {\enquote {\bibinfo {title} {{Scaling Ion Trap
  Quantum Computation through Fast Quantum Gates}},}\ }\href {\doibase
  10.1103/PhysRevLett.93.100502} {\bibfield  {journal} {\bibinfo  {journal}
  {Physical Review Letters}\ }\textbf {\bibinfo {volume} {93}},\ \bibinfo
  {pages} {100502} (\bibinfo {year} {2004})}\BibitemShut {NoStop}%
\bibitem [{\citenamefont {Taylor}\ \emph {et~al.}(2017)\citenamefont {Taylor},
  \citenamefont {Bentley}, \citenamefont {Pedernales}, \citenamefont {Lamata},
  \citenamefont {Solano}, \citenamefont {Carvalho},\ and\ \citenamefont
  {Hope}}]{Taylor2017}%
  \BibitemOpen
  \bibfield  {author} {\bibinfo {author} {\bibfnamefont {R.~L.}\ \bibnamefont
  {Taylor}}, \bibinfo {author} {\bibfnamefont {C.~D.~B.}\ \bibnamefont
  {Bentley}}, \bibinfo {author} {\bibfnamefont {J.~S.}\ \bibnamefont
  {Pedernales}}, \bibinfo {author} {\bibfnamefont {L.}~\bibnamefont {Lamata}},
  \bibinfo {author} {\bibfnamefont {E.}~\bibnamefont {Solano}}, \bibinfo
  {author} {\bibfnamefont {A.~R.~R.}\ \bibnamefont {Carvalho}}, \ and\ \bibinfo
  {author} {\bibfnamefont {J.~J.}\ \bibnamefont {Hope}},\ }\bibfield  {title}
  {\enquote {\bibinfo {title} {{A Study on Fast Gates for Large-Scale Quantum
  Simulation with Trapped Ions}},}\ }\href {\doibase 10.1038/srep46197}
  {\bibfield  {journal} {\bibinfo  {journal} {Scientific Reports}\ }\textbf
  {\bibinfo {volume} {7}},\ \bibinfo {pages} {46197} (\bibinfo {year}
  {2017})}\BibitemShut {NoStop}%
\bibitem [{\citenamefont {Bentley}\ \emph {et~al.}(2015)\citenamefont
  {Bentley}, \citenamefont {Carvalho},\ and\ \citenamefont
  {Hope}}]{Bentley2015}%
  \BibitemOpen
  \bibfield  {author} {\bibinfo {author} {\bibfnamefont {C.~D.~B.}\
  \bibnamefont {Bentley}}, \bibinfo {author} {\bibfnamefont {A.~R.~R.}\
  \bibnamefont {Carvalho}}, \ and\ \bibinfo {author} {\bibfnamefont {J.~J.}\
  \bibnamefont {Hope}},\ }\bibfield  {title} {\enquote {\bibinfo {title}
  {{Trapped ion scaling with pulsed fast gates}},}\ }\href {\doibase
  10.1088/1367-2630/17/10/103025} {\bibfield  {journal} {\bibinfo  {journal}
  {New Journal of Physics}\ }\textbf {\bibinfo {volume} {17}},\ \bibinfo
  {pages} {103025} (\bibinfo {year} {2015})}\BibitemShut {NoStop}%
\bibitem [{\citenamefont {Ratcliffe}\ \emph {et~al.}(2018)\citenamefont
  {Ratcliffe}, \citenamefont {Taylor}, \citenamefont {Hope},\ and\
  \citenamefont {Carvalho}}]{Ratcliffe2018}%
  \BibitemOpen
  \bibfield  {author} {\bibinfo {author} {\bibfnamefont {A.~K.}\ \bibnamefont
  {Ratcliffe}}, \bibinfo {author} {\bibfnamefont {R.~L.}\ \bibnamefont
  {Taylor}}, \bibinfo {author} {\bibfnamefont {J.~J.}\ \bibnamefont {Hope}}, \
  and\ \bibinfo {author} {\bibfnamefont {A.~R.~R.}\ \bibnamefont {Carvalho}},\
  }\bibfield  {title} {\enquote {\bibinfo {title} {{Scaling Trapped Ion Quantum
  Computers Using Fast Gates and Microtraps}},}\ }\href {\doibase
  10.1103/PhysRevLett.120.220501} {\bibfield  {journal} {\bibinfo  {journal}
  {Physical Review Letters}\ }\textbf {\bibinfo {volume} {120}},\ \bibinfo
  {pages} {220501} (\bibinfo {year} {2018})}\BibitemShut {NoStop}%
\bibitem [{\citenamefont {Johnson}\ \emph {et~al.}(2017)\citenamefont
  {Johnson}, \citenamefont {Wong-Campos}, \citenamefont {Neyenhuis},
  \citenamefont {Mizrahi},\ and\ \citenamefont {Monroe}}]{Johnson2017}%
  \BibitemOpen
  \bibfield  {author} {\bibinfo {author} {\bibfnamefont {K.~G.}\ \bibnamefont
  {Johnson}}, \bibinfo {author} {\bibfnamefont {J.~D.}\ \bibnamefont
  {Wong-Campos}}, \bibinfo {author} {\bibfnamefont {B.}~\bibnamefont
  {Neyenhuis}}, \bibinfo {author} {\bibfnamefont {J.}~\bibnamefont {Mizrahi}},
  \ and\ \bibinfo {author} {\bibfnamefont {C.}~\bibnamefont {Monroe}},\
  }\bibfield  {title} {\enquote {\bibinfo {title} {{Ultrafast creation of large
  Schr{\"{o}}dinger cat states of an atom}},}\ }\href {\doibase
  10.1038/s41467-017-00682-6} {\bibfield  {journal} {\bibinfo  {journal}
  {Nature Communications}\ }\textbf {\bibinfo {volume} {8}},\ \bibinfo {pages}
  {697} (\bibinfo {year} {2017})}\BibitemShut {NoStop}%
\bibitem [{\citenamefont {Wong-Campos}\ \emph {et~al.}(2017)\citenamefont
  {Wong-Campos}, \citenamefont {Moses}, \citenamefont {Johnson},\ and\
  \citenamefont {Monroe}}]{Wong-Campos2017}%
  \BibitemOpen
  \bibfield  {author} {\bibinfo {author} {\bibfnamefont {J.~D.}\ \bibnamefont
  {Wong-Campos}}, \bibinfo {author} {\bibfnamefont {S.~A.}\ \bibnamefont
  {Moses}}, \bibinfo {author} {\bibfnamefont {K.~G.}\ \bibnamefont {Johnson}},
  \ and\ \bibinfo {author} {\bibfnamefont {C.}~\bibnamefont {Monroe}},\
  }\bibfield  {title} {\enquote {\bibinfo {title} {{Demonstration of Two-Atom
  Entanglement with Ultrafast Optical Pulses}},}\ }\href {\doibase
  10.1103/PhysRevLett.119.230501} {\bibfield  {journal} {\bibinfo  {journal}
  {Physical Review Letters}\ }\textbf {\bibinfo {volume} {119}},\ \bibinfo
  {pages} {230501} (\bibinfo {year} {2017})}\BibitemShut {NoStop}%
\bibitem [{\citenamefont {Sch{\"{a}}fer}\ \emph {et~al.}(2018)\citenamefont
  {Sch{\"{a}}fer}, \citenamefont {Ballance}, \citenamefont {Thirumalai},
  \citenamefont {Stephenson}, \citenamefont {Ballance}, \citenamefont
  {Steane},\ and\ \citenamefont {Lucas}}]{Schafer2017}%
  \BibitemOpen
  \bibfield  {author} {\bibinfo {author} {\bibfnamefont {V.~M.}\ \bibnamefont
  {Sch{\"{a}}fer}}, \bibinfo {author} {\bibfnamefont {C.~J.}\ \bibnamefont
  {Ballance}}, \bibinfo {author} {\bibfnamefont {K.}~\bibnamefont
  {Thirumalai}}, \bibinfo {author} {\bibfnamefont {L.~J.}\ \bibnamefont
  {Stephenson}}, \bibinfo {author} {\bibfnamefont {T.~G.}\ \bibnamefont
  {Ballance}}, \bibinfo {author} {\bibfnamefont {A.~M.}\ \bibnamefont
  {Steane}}, \ and\ \bibinfo {author} {\bibfnamefont {D.~M.}\ \bibnamefont
  {Lucas}},\ }\bibfield  {title} {\enquote {\bibinfo {title} {{Fast quantum
  logic gates with trapped-ion qubits}},}\ }\href {\doibase
  10.1038/nature25737} {\bibfield  {journal} {\bibinfo  {journal} {Nature}\
  }\textbf {\bibinfo {volume} {555}},\ \bibinfo {pages} {75} (\bibinfo {year}
  {2018})}\BibitemShut {NoStop}%
\bibitem [{\citenamefont {Shimshoni}\ \emph {et~al.}(2011)\citenamefont
  {Shimshoni}, \citenamefont {Morigi},\ and\ \citenamefont
  {Fishman}}]{Shimshoni2011}%
  \BibitemOpen
  \bibfield  {author} {\bibinfo {author} {\bibfnamefont {E.}~\bibnamefont
  {Shimshoni}}, \bibinfo {author} {\bibfnamefont {G.}~\bibnamefont {Morigi}}, \
  and\ \bibinfo {author} {\bibfnamefont {S.}~\bibnamefont {Fishman}},\
  }\bibfield  {title} {\enquote {\bibinfo {title} {{Quantum Zigzag Transition
  in Ion Chains}},}\ }\href {\doibase 10.1103/PhysRevLett.106.010401}
  {\bibfield  {journal} {\bibinfo  {journal} {Physical Review Letters}\
  }\textbf {\bibinfo {volume} {106}},\ \bibinfo {pages} {010401} (\bibinfo
  {year} {2011})}\BibitemShut {NoStop}%
\bibitem [{\citenamefont {Leibfried}\ \emph {et~al.}(2003)\citenamefont
  {Leibfried}, \citenamefont {Blatt}, \citenamefont {Monroe},\ and\
  \citenamefont {Wineland}}]{Leibfried2003}%
  \BibitemOpen
  \bibfield  {author} {\bibinfo {author} {\bibfnamefont {D.}~\bibnamefont
  {Leibfried}}, \bibinfo {author} {\bibfnamefont {R.}~\bibnamefont {Blatt}},
  \bibinfo {author} {\bibfnamefont {C.}~\bibnamefont {Monroe}}, \ and\ \bibinfo
  {author} {\bibfnamefont {D.}~\bibnamefont {Wineland}},\ }\bibfield  {title}
  {\enquote {\bibinfo {title} {{Quantum dynamics of single trapped ions}},}\
  }\href {\doibase 10.1103/RevModPhys.75.281} {\bibfield  {journal} {\bibinfo
  {journal} {Reviews of Modern Physics}\ }\textbf {\bibinfo {volume} {75}},\
  \bibinfo {pages} {281} (\bibinfo {year} {2003})}\BibitemShut {NoStop}%
\bibitem [{\citenamefont {Shen}\ and\ \citenamefont {Duan}(2014)}]{Shen2014a}%
  \BibitemOpen
  \bibfield  {author} {\bibinfo {author} {\bibfnamefont {C.}~\bibnamefont
  {Shen}}\ and\ \bibinfo {author} {\bibfnamefont {L.-M.}\ \bibnamefont
  {Duan}},\ }\bibfield  {title} {\enquote {\bibinfo {title} {{High-fidelity
  quantum gates for trapped ions under micromotion}},}\ }\href {\doibase
  10.1103/PhysRevA.90.022332} {\bibfield  {journal} {\bibinfo  {journal}
  {Physical Review A}\ }\textbf {\bibinfo {volume} {90}},\ \bibinfo {pages}
  {022332} (\bibinfo {year} {2014})}\BibitemShut {NoStop}%
\bibitem [{\citenamefont {Bermudez}\ \emph {et~al.}(2017)\citenamefont
  {Bermudez}, \citenamefont {Schindler}, \citenamefont {Monz}, \citenamefont
  {Blatt},\ and\ \citenamefont {M{\"{u}}ller}}]{Bermudez2017}%
  \BibitemOpen
  \bibfield  {author} {\bibinfo {author} {\bibfnamefont {A.}~\bibnamefont
  {Bermudez}}, \bibinfo {author} {\bibfnamefont {P.}~\bibnamefont {Schindler}},
  \bibinfo {author} {\bibfnamefont {T.}~\bibnamefont {Monz}}, \bibinfo {author}
  {\bibfnamefont {R.}~\bibnamefont {Blatt}}, \ and\ \bibinfo {author}
  {\bibfnamefont {M.}~\bibnamefont {M{\"{u}}ller}},\ }\bibfield  {title}
  {\enquote {\bibinfo {title} {{Micromotion-enabled improvement of quantum
  logic gates with trapped ions}},}\ }\href {\doibase 10.1088/1367-2630/aa86eb}
  {\bibfield  {journal} {\bibinfo  {journal} {New Journal of Physics}\ }\textbf
  {\bibinfo {volume} {19}},\ \bibinfo {pages} {113038} (\bibinfo {year}
  {2017})}\BibitemShut {NoStop}%
\bibitem [{\citenamefont {Kumph}\ \emph {et~al.}(2016)\citenamefont {Kumph},
  \citenamefont {Holz}, \citenamefont {Langer}, \citenamefont {Meraner},
  \citenamefont {Niedermayr}, \citenamefont {Brownnutt},\ and\ \citenamefont
  {Blatt}}]{Kumph2016}%
  \BibitemOpen
  \bibfield  {author} {\bibinfo {author} {\bibfnamefont {M.}~\bibnamefont
  {Kumph}}, \bibinfo {author} {\bibfnamefont {P.}~\bibnamefont {Holz}},
  \bibinfo {author} {\bibfnamefont {K.}~\bibnamefont {Langer}}, \bibinfo
  {author} {\bibfnamefont {M.}~\bibnamefont {Meraner}}, \bibinfo {author}
  {\bibfnamefont {M.}~\bibnamefont {Niedermayr}}, \bibinfo {author}
  {\bibfnamefont {M.}~\bibnamefont {Brownnutt}}, \ and\ \bibinfo {author}
  {\bibfnamefont {R.}~\bibnamefont {Blatt}},\ }\bibfield  {title} {\enquote
  {\bibinfo {title} {{Operation of a planar-electrode ion-trap array with
  adjustable RF electrodes}},}\ }\href {\doibase 10.1088/1367-2630/18/2/023047}
  {\bibfield  {journal} {\bibinfo  {journal} {New Journal of Physics}\ }\textbf
  {\bibinfo {volume} {18}},\ \bibinfo {pages} {023047} (\bibinfo {year}
  {2016})}\BibitemShut {NoStop}%
\bibitem [{\citenamefont {Landa}\ \emph {et~al.}(2012)\citenamefont {Landa},
  \citenamefont {Drewsen}, \citenamefont {Reznik},\ and\ \citenamefont
  {Retzker}}]{Landa2012}%
  \BibitemOpen
  \bibfield  {author} {\bibinfo {author} {\bibfnamefont {H.}~\bibnamefont
  {Landa}}, \bibinfo {author} {\bibfnamefont {M.}~\bibnamefont {Drewsen}},
  \bibinfo {author} {\bibfnamefont {B.}~\bibnamefont {Reznik}}, \ and\ \bibinfo
  {author} {\bibfnamefont {A.}~\bibnamefont {Retzker}},\ }\bibfield  {title}
  {\enquote {\bibinfo {title} {{Modes of oscillation in radiofrequency Paul
  traps}},}\ }\href {\doibase 10.1088/1367-2630/14/9/093023} {\bibfield
  {journal} {\bibinfo  {journal} {New Journal of Physics}\ }\textbf {\bibinfo
  {volume} {14}},\ \bibinfo {pages} {93023} (\bibinfo {year}
  {2012})}\BibitemShut {NoStop}%
\bibitem [{\citenamefont {Garcia-Ripoll}\ \emph {et~al.}(2003)\citenamefont
  {Garcia-Ripoll}, \citenamefont {Zoller},\ and\ \citenamefont
  {Cirac}}]{Garcia-Ripoll2003}%
  \BibitemOpen
  \bibfield  {author} {\bibinfo {author} {\bibfnamefont {J.~J.}\ \bibnamefont
  {Garcia-Ripoll}}, \bibinfo {author} {\bibfnamefont {P.}~\bibnamefont
  {Zoller}}, \ and\ \bibinfo {author} {\bibfnamefont {J.~I.}\ \bibnamefont
  {Cirac}},\ }\bibfield  {title} {\enquote {\bibinfo {title} {{Fast and robust
  two-qubit gates for scalable ion trap quantum computing}},}\ }\href {\doibase
  10.1103/PhysRevLett.91.157901} {\bibfield  {journal} {\bibinfo  {journal}
  {Physical Review Letters}\ }\textbf {\bibinfo {volume} {91}},\ \bibinfo
  {pages} {157901} (\bibinfo {year} {2003})}\BibitemShut {NoStop}%
\bibitem [{\citenamefont {Bentley}\ \emph {et~al.}(2016)\citenamefont
  {Bentley}, \citenamefont {Taylor}, \citenamefont {Carvalho},\ and\
  \citenamefont {Hope}}]{Bentley2016}%
  \BibitemOpen
  \bibfield  {author} {\bibinfo {author} {\bibfnamefont {C.~D.~B.}\
  \bibnamefont {Bentley}}, \bibinfo {author} {\bibfnamefont {R.~L.}\
  \bibnamefont {Taylor}}, \bibinfo {author} {\bibfnamefont {A.~R.~R.}\
  \bibnamefont {Carvalho}}, \ and\ \bibinfo {author} {\bibfnamefont {J.~J.}\
  \bibnamefont {Hope}},\ }\bibfield  {title} {\enquote {\bibinfo {title}
  {{Stability thresholds and calculation techniques for fast entangling gates
  on trapped ions}},}\ }\href {\doibase 10.1103/PhysRevA.93.042342} {\bibfield
  {journal} {\bibinfo  {journal} {Physical Review A}\ }\textbf {\bibinfo
  {volume} {93}},\ \bibinfo {pages} {042342} (\bibinfo {year}
  {2016})}\BibitemShut {NoStop}%
\bibitem [{\citenamefont {Inlek}\ \emph {et~al.}(2017)\citenamefont {Inlek},
  \citenamefont {Crocker}, \citenamefont {Lichtman}, \citenamefont {Sosnova},\
  and\ \citenamefont {Monroe}}]{Inlek2017}%
  \BibitemOpen
  \bibfield  {author} {\bibinfo {author} {\bibfnamefont {I.~V.}\ \bibnamefont
  {Inlek}}, \bibinfo {author} {\bibfnamefont {C.}~\bibnamefont {Crocker}},
  \bibinfo {author} {\bibfnamefont {M.}~\bibnamefont {Lichtman}}, \bibinfo
  {author} {\bibfnamefont {K.}~\bibnamefont {Sosnova}}, \ and\ \bibinfo
  {author} {\bibfnamefont {C.}~\bibnamefont {Monroe}},\ }\bibfield  {title}
  {\enquote {\bibinfo {title} {{Multispecies Trapped-Ion Node for Quantum
  Networking}},}\ }\href {\doibase 10.1103/PhysRevLett.118.250502} {\bibfield
  {journal} {\bibinfo  {journal} {Physical Review Letters}\ }\textbf {\bibinfo
  {volume} {118}},\ \bibinfo {pages} {250502} (\bibinfo {year}
  {2017})}\BibitemShut {NoStop}%
\bibitem [{\citenamefont {Drakoudis}\ \emph {et~al.}(2006)\citenamefont
  {Drakoudis}, \citenamefont {S{\"{o}}llner},\ and\ \citenamefont
  {Werth}}]{Drakoudis2006}%
  \BibitemOpen
  \bibfield  {author} {\bibinfo {author} {\bibfnamefont {A.}~\bibnamefont
  {Drakoudis}}, \bibinfo {author} {\bibfnamefont {M.}~\bibnamefont
  {S{\"{o}}llner}}, \ and\ \bibinfo {author} {\bibfnamefont {G.}~\bibnamefont
  {Werth}},\ }\bibfield  {title} {\enquote {\bibinfo {title} {{Instabilities of
  ion motion in a linear Paul trap}},}\ }\href {\doibase
  10.1016/j.ijms.2006.02.006} {\bibfield  {journal} {\bibinfo  {journal}
  {International Journal of Mass Spectrometry}\ }\textbf {\bibinfo {volume}
  {252}},\ \bibinfo {pages} {61} (\bibinfo {year} {2006})}\BibitemShut
  {NoStop}%
\bibitem [{\citenamefont {Alheit}\ \emph {et~al.}(1996)\citenamefont {Alheit},
  \citenamefont {Gudjons}, \citenamefont {Kleineidam},\ and\ \citenamefont
  {Werth}}]{Alheit1996}%
  \BibitemOpen
  \bibfield  {author} {\bibinfo {author} {\bibfnamefont {R.}~\bibnamefont
  {Alheit}}, \bibinfo {author} {\bibfnamefont {T.}~\bibnamefont {Gudjons}},
  \bibinfo {author} {\bibfnamefont {S.}~\bibnamefont {Kleineidam}}, \ and\
  \bibinfo {author} {\bibfnamefont {G.}~\bibnamefont {Werth}},\ }\bibfield
  {title} {\enquote {\bibinfo {title} {{Some Observations on Higher-order
  Non-linear Resonances in a Paul Trap}},}\ }\href {\doibase
  10.1002/(SICI)1097-0231(19960331)10:5<583::AID-RCM497>3.3.CO;2-U} {\bibfield
  {journal} {\bibinfo  {journal} {Rapid Communications in Mass Spectrometry}\
  }\textbf {\bibinfo {volume} {10}},\ \bibinfo {pages} {583} (\bibinfo {year}
  {1996})}\BibitemShut {NoStop}%
\bibitem [{\citenamefont {Collings}\ and\ \citenamefont
  {Douglas}(2000)}]{Collings2000}%
  \BibitemOpen
  \bibfield  {author} {\bibinfo {author} {\bibfnamefont {B.~A.}\ \bibnamefont
  {Collings}}\ and\ \bibinfo {author} {\bibfnamefont {D.~J.}\ \bibnamefont
  {Douglas}},\ }\bibfield  {title} {\enquote {\bibinfo {title} {{Observation of
  higher order quadrupole excitation frequencies in a linear ion trap}},}\
  }\href {\doibase 10.1016/S1044-0305(00)00171-9} {\bibfield  {journal}
  {\bibinfo  {journal} {Journal of the American Society for Mass Spectrometry}\
  }\textbf {\bibinfo {volume} {11}},\ \bibinfo {pages} {1016} (\bibinfo {year}
  {2000})}\BibitemShut {NoStop}%
\bibitem [{\citenamefont {Bentley}\ \emph {et~al.}(2013)\citenamefont
  {Bentley}, \citenamefont {Carvalho}, \citenamefont {Kielpinski},\ and\
  \citenamefont {Hope}}]{Bentley2012a}%
  \BibitemOpen
  \bibfield  {author} {\bibinfo {author} {\bibfnamefont {C.~D.~B.}\
  \bibnamefont {Bentley}}, \bibinfo {author} {\bibfnamefont {A.~R.~R.}\
  \bibnamefont {Carvalho}}, \bibinfo {author} {\bibfnamefont {D.}~\bibnamefont
  {Kielpinski}}, \ and\ \bibinfo {author} {\bibfnamefont {J.~J.}\ \bibnamefont
  {Hope}},\ }\bibfield  {title} {\enquote {\bibinfo {title} {{Fast gates for
  ion traps by splitting laser pulses}},}\ }\href {\doibase
  10.1088/1367-2630/15/4/043006} {\bibfield  {journal} {\bibinfo  {journal}
  {New Journal of Physics}\ }\textbf {\bibinfo {volume} {15}},\ \bibinfo
  {pages} {043006} (\bibinfo {year} {2013})}\BibitemShut {NoStop}%
\bibitem [{\citenamefont {Heinrich}\ \emph {et~al.}(2019)\citenamefont
  {Heinrich}, \citenamefont {Guggemos}, \citenamefont {Guevara-Bertsch},
  \citenamefont {Hussain}, \citenamefont {Roos},\ and\ \citenamefont
  {Blatt}}]{Heinrich2019}%
  \BibitemOpen
  \bibfield  {author} {\bibinfo {author} {\bibfnamefont {D.}~\bibnamefont
  {Heinrich}}, \bibinfo {author} {\bibfnamefont {M.}~\bibnamefont {Guggemos}},
  \bibinfo {author} {\bibfnamefont {M.}~\bibnamefont {Guevara-Bertsch}},
  \bibinfo {author} {\bibfnamefont {M.~I.}\ \bibnamefont {Hussain}}, \bibinfo
  {author} {\bibfnamefont {C.~F.}\ \bibnamefont {Roos}}, \ and\ \bibinfo
  {author} {\bibfnamefont {R.}~\bibnamefont {Blatt}},\ }\bibfield  {title}
  {\enquote {\bibinfo {title} {{Ultrafast coherent excitation of a
  {\$}{\^{}}{\{}40{\}}{\$}Ca{\$}{\^{}}+{\$} ion}},}\ }\href {\doibase
  10.1088/1367-2630/ab2a7e} {\bibfield  {journal} {\bibinfo  {journal} {New
  Journal of Physics}\ }\textbf {\bibinfo {volume} {21}},\ \bibinfo {pages}
  {073017} (\bibinfo {year} {2019})}\BibitemShut {NoStop}%
\bibitem [{\citenamefont {Gale}\ \emph {et~al.}(2019)\citenamefont {Gale},
  \citenamefont {Mehdi}, \citenamefont {Oberg}, \citenamefont {Ratcliffe},
  \citenamefont {Haine},\ and\ \citenamefont {Hope}}]{Gale2019}%
  \BibitemOpen
  \bibfield  {author} {\bibinfo {author} {\bibfnamefont {E.~P.~G.}\
  \bibnamefont {Gale}}, \bibinfo {author} {\bibfnamefont {Z.}~\bibnamefont
  {Mehdi}}, \bibinfo {author} {\bibfnamefont {L.~M.}\ \bibnamefont {Oberg}},
  \bibinfo {author} {\bibfnamefont {A.~K.}\ \bibnamefont {Ratcliffe}}, \bibinfo
  {author} {\bibfnamefont {S.~A.}\ \bibnamefont {Haine}}, \ and\ \bibinfo
  {author} {\bibfnamefont {J.~J.}\ \bibnamefont {Hope}},\ }\bibfield  {title}
  {\enquote {\bibinfo {title} {{Optimised fast gates for quantum computing with
  trapped ions}},}\ }\href {http://arxiv.org/abs/1912.07780} {\bibfield
  {journal} {\bibinfo  {journal} {arXiv preprint}\ } (\bibinfo {year}
  {2019})}\BibitemShut {NoStop}%
\bibitem [{\citenamefont {Kabytayev}\ \emph {et~al.}(2014)\citenamefont
  {Kabytayev}, \citenamefont {Green}, \citenamefont {Khodjasteh}, \citenamefont
  {Biercuk}, \citenamefont {Viola},\ and\ \citenamefont
  {Brown}}]{Kabytayev2014}%
  \BibitemOpen
  \bibfield  {author} {\bibinfo {author} {\bibfnamefont {C.}~\bibnamefont
  {Kabytayev}}, \bibinfo {author} {\bibfnamefont {T.~J.}\ \bibnamefont
  {Green}}, \bibinfo {author} {\bibfnamefont {K.}~\bibnamefont {Khodjasteh}},
  \bibinfo {author} {\bibfnamefont {M.~J.}\ \bibnamefont {Biercuk}}, \bibinfo
  {author} {\bibfnamefont {L.}~\bibnamefont {Viola}}, \ and\ \bibinfo {author}
  {\bibfnamefont {K.~R.}\ \bibnamefont {Brown}},\ }\bibfield  {title} {\enquote
  {\bibinfo {title} {{Robustness of composite pulses to time-dependent control
  noise}},}\ }\href {\doibase 10.1103/PhysRevA.90.012316} {\bibfield  {journal}
  {\bibinfo  {journal} {Physical Review A}\ }\textbf {\bibinfo {volume} {90}},\
  \bibinfo {pages} {012316} (\bibinfo {year} {2014})}\BibitemShut {NoStop}%
\bibitem [{\citenamefont {Vandersypen}\ and\ \citenamefont
  {Chuang}(2005)}]{Vandersypen2005}%
  \BibitemOpen
  \bibfield  {author} {\bibinfo {author} {\bibfnamefont {L.~M.~K.}\
  \bibnamefont {Vandersypen}}\ and\ \bibinfo {author} {\bibfnamefont {I.~L.}\
  \bibnamefont {Chuang}},\ }\bibfield  {title} {\enquote {\bibinfo {title}
  {{NMR techniques for quantum control and computation}},}\ }\href {\doibase
  10.1103/RevModPhys.76.1037} {\bibfield  {journal} {\bibinfo  {journal}
  {Reviews of Modern Physics}\ }\textbf {\bibinfo {volume} {76}},\ \bibinfo
  {pages} {1037} (\bibinfo {year} {2005})}\BibitemShut {NoStop}%
\end{thebibliography}%
\end{document}